\documentclass[11pt,a4paper]{article}
\pdfoutput=1
\usepackage{jheppub}

\usepackage{ifpdf}
\usepackage{afterpage}
\usepackage{amssymb}
\usepackage{amsmath}
\usepackage{graphicx}
\usepackage{longtable}
\usepackage{verbatim}
\usepackage{amsfonts}
\usepackage{bbold}

\usepackage[table,usenames,dvipsnames]{xcolor}
\usepackage{titlesec}
\usepackage{enumerate}
\usepackage{float}
\usepackage[multiple]{footmisc}
\newcommand\numberthis{\addtocounter{equation}{1}\tag{\theequation}}

\usepackage{physics}
\usepackage{subcaption}

\usepackage{tikz}
\usetikzlibrary{trees}
\usetikzlibrary{decorations.pathmorphing}
\usetikzlibrary{decorations.markings}
\usetikzlibrary{decorations.pathreplacing}
\tikzset{
    photon/.style={decorate, decoration={snake}, draw=black},
    wino/.style={draw=redwine},
    electron/.style={draw=black, postaction={decorate},
        decoration={markings,mark=at position .55 with {\arrow[draw=black]{>}}}},
    scalar/.style={draw=black, dashed,postaction={decorate},
        decoration={markings,mark=at position .55 with {\arrow[draw=black]{>}}}},
    gluon/.style={decorate, draw=black,
        decoration={coil,amplitude=4pt, segment length=5pt}}
}

\newcommand{\bear}{\begin{array}}
\newcommand{\ear}{\end{array}}
\newcommand{\beq}{\begin{equation}}
\newcommand{\eeq}{\end{equation}}
\newcommand{\beqa}{\begin{eqnarray}}
\newcommand{\eeqa}{\end{eqnarray}}

\def\OMIT#1{{}}
\newcommand{\lsim}{\mathrel{\rlap{\lower4pt\hbox{\hskip1pt$\sim$}}
    \raise1pt\hbox{$<$}}}         %less than or approx. symbol
\newcommand{\gsim}{\mathrel{\rlap{\lower4pt\hbox{\hskip1pt$\sim$}}
    \raise1pt\hbox{$>$}}}         %greater than or approx. symbol
\newcommand{\Fig}[1]{Fig.~\ref{#1}}

\newcommand{\Eq}[1]{Eq.~(\ref{#1})}

\newcommand{\ignore}[1]{}

\vspace*{-30mm}

\title{\boldmath Phenomenology of Spillway Preheating: Equation of State and Gravitational Waves}
\author[a]{Gareth Mansfield,}
\author[a,b]{JiJi Fan,}
\author[c, d]{Qianshu Lu}
\affiliation[a]{Department of Physics, Brown University,
Providence, RI, 02912, USA}
\affiliation[b]{Brown Theoretical Physics Center, Brown University, Providence, RI, 02912, USA}
\affiliation[c]{School of Natural Sciences, Institute for Advanced Study, Princeton, NJ 08540, USA}
\affiliation[d]{Center for Cosmology and Particle Physics, Department of Physics, New York University, New York, NY 10003, USA}
\emailAdd{gareth\_mansfield@brown.edu}
\emailAdd{jiji\_fan@brown.edu}
\emailAdd{qianshu.lu@ias.edu}

\vspace*{1cm}

\abstract{In the canonical tachyonic resonance preheating scenario, only an order one fraction of energy density in the inflaton is transferred to radiation, due to backreaction effects. One possible way to improve the energy transfer efficiency is to allow for the perturbative decays of the resonantly produced daughter particles, which serve as the ``spillway" to drain the direct decay products from inflaton and to reduce the backreaction. In this article, we study two observational consequences of spillway preheating. The first is on the inflationary observables: the scalar spectrum tilt $n_s$ and tensor-to-scalar ratio $r$. The spillway scenario modifies the evolution of the equation of state between the end of inflation and the thermal big bang. As a result, it affects the time elapsed from inflation to the Cosmic Microwave Background (CMB), as well as the fits of inflationary models and their corresponding prediction for $n_s$ and $r$. We map out the equation of state by systematically scanning the parameter space of the spillway scenario, and show that the most efficient spillway scenario predicts a bluer spectrum, compared to the tachyonic preheating scenario. Another consequence is the production of high-frequency gravitational waves (GWs). Comparing the simulation results with those of tachyonic preheating, we find that the existence of spillways leads to sharper-peaked GW spectra with a mildly damped amplitude.      }

\begin{document}

\maketitle

%%%%%%%%%%%%%%%%%%%%%%%%%%%%%%%%%%%%%%%%%
\section{Introduction}
\label{sec:intro}
%%%%%%%%%%%%%%%%%%%%%%%%%%%%%%%%%%%%%%%%%

So far, cosmological observations have provided compelling evidence for an exponentially expanding inflationary phase in the early Universe and a hot big bang shortly after it. Yet the intermediate stage between the two remains mysterious and is often referred to as the ``primordial dark age", simply reflecting our ignorance of this connecting phase. It has been generally assumed that the phase transition is achieved through (p)reheating processes converting the inflaton energy to the thermal energies of other particles. More specifically, this conversion could be through either perturbative decays of the inflaton~\cite{Abbott:1982hn, Dolgov:1982th, Albrecht:1982mp}, the so-called reheating, or non-perturbative and out-of-equilibrium dynamics~\cite{Traschen:1990sw,Dolgov:1989us, Shtanov:1994ce, Kofman:1994rk, Boyanovsky:1995ud, Yoshimura:1995gc, Kaiser:1995fb, Kofman:1997yn,Allahverdi:2010xz,Amin:2014eta}. The latter processes are referred to as preheating, usually happening much faster and earlier than the reheating ones. 

Preheating contains a plethora of rich dynamics beyond the reach of perturbative calculations, and requires a better understanding. Yet the canonical most-studied preheating scenarios, such as parametric resonance~\cite{Kofman:1997yn} and tachyonic resonance~\cite{Dufaux:2006ee}, could only transfer at most an order one fraction of the inflaton energy to radiation.\footnote{The only known exception in the literature is the tachyonic gauge preheating by coupling the (pseudo-)scalar inflaton to the (dual) gauge field strength~\cite{Deskins:2013dwa,Adshead:2015pva, Adshead:2017xll, Cuissa:2018oiw}, which can boost the depletion of the inflaton energy density by up to two orders of magnitude.} Reheating is needed to complete the transition to the thermal big bang at a (much) later stage. This raises an intriguing question of whether there exists a more efficient preheating mechanism to deplete more inflaton energy and convert it into radiation. 

One such possibility is the ``spillway" preheating, which could improve the depletion of the inflaton energy density by up to four orders of magnitude~\cite{Fan:2021otj}. The bottleneck of traditional preheating mechanisms is that although initially the daughter particles, e.g., scalars denoted by $\chi$'s, could be produced copiously through exponential non-perturbative processes, they would backreact on the inflaton, $\phi$, pause the production processes, and prevent further energy transfer. For example, the standard tachyonic resonance production could only transfer about half of the inflaton energy to radiation~\cite{Dufaux:2006ee}. To overcome this difficulty and reduce the backreaction, a ``spillway" is introduced through the perturbative decays of $\chi$'s to second-generation daughter fermions $\psi$ via a Yukawa coupling. The cascade decays $\phi \to \chi \to \psi$, combining non-perturbative decays as the first step and perturbative decays as the second, is demonstrated to enhance the inflaton energy transfer significantly in some parameter space that could be simulated numerically~\cite{Fan:2021otj}.\footnote{Comparison of the spillway mechanism with some similar earlier studies relying on multi-step decays~\cite{Felder:1998vq, Garcia-Bellido:2008ycs, Repond:2016sol} could be found in~\cite{Fan:2021otj}. Other aspects on the interplay between non-perturbative and perturbative processes after inflation have been studied in~\cite{Kasuya:1996np, Bezrukov:2008ut,Mukaida:2012bz,Kost:2021rbi,Garcia:2021iag}. } 

In this paper, we examine two potential observational consequences of spillway preheating and compare them with those of the tachyonic resonance scenario. The first one we study is the impact on two inflationary observables: scalar spectrum index $n_s$ and tensor-to-scalar ratio $r$ (earlier studies of the (p)reheating impact on these observables could be found in~\cite{Liddle:2003as,Dai:2014jja,Munoz:2014eqa,Martin:2016oyk,Hardwick:2016whe,Lozanov:2016hid,Lozanov:2017hjm,Antusch:2020iyq, Bettoni:2021zhq, Antusch:2022mqv,Lodman:2023yrc, Barman:2023opy}). The spillway scenario could modify the evolution of the equation of state in the cosmic dark age, and affect the time elapsed from inflation to the Cosmic Microwave Background (CMB). As a result, this could influence the fits of inflationary models and their corresponding prediction for $n_s$ and $r$. We have implemented a comprehensive scan of the equation of state in the input model parameter space, which has not been done before. We then discuss the effects on the fits to $n_s$ and $r$ in different classes of inflationary models. The next observable we explore is the generation of high-frequency gravitational waves (GWs). The non-linear dynamics leads to fragmentation of the inflaton field and an inhomogeneous matter distribution, sourcing GWs. GWs from canonical preheating models have been studied in~\cite{Khlebnikov:1997di,Easther:2006vd,Easther:2006gt,GarciaBellido:2007af,Dufaux:2007pt,Dufaux:2008dn,Dufaux:2010cf,Bethke:2013vca,Adshead:2018doq,Kitajima:2018zco,Bartolo:2016ami,Figueroa:2017vfa,Caprini:2018mtu,Bartolo:2018qqn,Lozanov:2019ylm,Adshead:2019igv,Adshead:2019lbr}. We will present simulation results of GWs for the spillway preheating scenario, which share some common properties with those of the tachyonic resonance scenario, but also possess their own distinctive feature.

The paper is organized as follows: in Sec~\ref{sec:model}, we review the basics of the tachyonic resonance and spillway preheating mechanisms. In Sec.~\ref{sec:obs1}, we scan the parameter space, and present how the equation of state evolves in spillway scenario, comparing it with that in the tachyonic case. We then discuss how it affects fits of $n_s$ and $r$ in various inflationary models. In Sec.~\ref{sec:obs2}, we show how the GW production depends on the model parameters and compare the GWs produced in the spillway and tachyonic preheating scenarios. We conclude in Sec.~\ref{sec:conclusions}.

%%%%%%%%%%%%%%%%%%%%%%%%%%%%%%%%%%%%%%%%%
\section{Models}
\label{sec:model}
%%%%%%%%%%%%%%%%%%%%%%%%%%%%%%%%%%%%%%%%%
In this section, we outline two efficient preheating models of interest which rely on non-perturbative particle production. We first discuss key features of the tachyonic resonance model introduced in~\cite{Dufaux:2006ee}. We then review its variant, spillway preheating studied in~\cite{Fan:2021otj}, and emphasize the main differences between the two scenarios as well as the advantage of spillway preheating over the canonical tachyonic resonance.

%%%%%%%%%%%%%%%%%%%%%%%%%%%%%%%%%%%%%%%%%
\subsection{Tachyonic Resonance Preheating}
\label{subsec:model1}
%%%%%%%%%%%%%%%%%%%%%%%%%%%%%%%%%%%%%%%%%

The simplest tachyonic resonance preheating model consists of a real inflaton field $\phi$ and a real scalar daughter particle $\chi$. It is governed by the potential
\begin{align}
    \label{eq:tachpotential}
    V (\phi, \chi)=\frac12 m^2\phi^2 +\frac12 \frac{M^2}f\phi\chi^2 +\frac14\lambda\chi^4~.
\end{align}
The energy scales of this model include the inflaton mass $m$, which we will fix to be $10^{-6}M_{\text{pl}}$ with the reduced Planck scale $M_{\text{pl}}\approx 2.4\times 10^{18} \, \text{GeV}$. At the end of inflation (with the time set at $t=0$), the inflaton field  starts to oscillate around the minimum of its potential with an initial amplitude $\phi_0=f$. Without loss of generality and for convenience, we ignore a possible quadratic mass term for $\chi$ in the potential. Yet the trilinear interaction between $\phi$ and $\chi$ still gives rise to an effective mass squared for $\chi$ equal to $\frac{M^2}{f}\phi$. When $\phi$ is on the positive side, the effective mass squared of $\chi$ is of order $M^2$ when the inflaton just starts to oscillate. When $\phi$ dips into the negative region, this effective mass squared becomes negative, triggering a tachyonic instability. On the branch of negative $\phi$, $\chi=0$ sits at an unstable equilibrium point of the potential, and quickly begins to grow after spontaneous symmetry breaking. This instability drives $\phi\to\chi\chi$ decays at a rate governed by the parameter~\cite{Amin:2018kkg}
\begin{align}
    \label{eq:q}
    q\equiv\frac{M^2}{m^2}~.
\end{align}
Higher values of $q$ correspond to more efficient particle production.

To ensure that the potential is bounded from below to prevent runaway production of $\chi$ particles, the tachyonic model requires a quartic self-interaction of $\chi$, $\lambda\chi^4$, with $\lambda$ being a positive dimensionless constant. This interaction leads to a positive contribution $\lambda\expval{\chi^2}$ to the effective mass squared of $\chi$, once the particle production starts. It competes against the tachyonic contribution from the trilinear coupling on the negative side of $\phi$, manifesting as a backreaction to slow down $\phi\to\chi\chi$ decays. We characterize the strength of this effect via the backreaction efficiency parameter, defined to be a product of the ratios between the energy in the trilinear interaction to the energy in the two fields~\cite{Amin:2018kkg}:
\begin{align}
    b \equiv \frac{1}{4}\left(\frac{\frac{1}{2} \frac{M^2}{f} \phi \chi^2}{\frac{1}{2} m^2 \phi^2}\right)\left(\frac{\frac{1}{2} \frac{M^2}{f} \phi \chi^2}{\frac{1}{4} \lambda \chi^4}\right)=\frac{M^4}{2 \lambda m^2 f^2}~.
\end{align}
The model requires $b\in[0,1)$, since setting $b\geq 1$ causes the potential to be unbounded from below. At early times, $m_\chi^2$ will be negative whenever $\expval{\phi}<0$, driving production of $\chi$ until the backreaction from $\expval{\chi^2}$ becomes large enough to win out against the tachyonic resonance. We can compute the critical 
value of $\expval{\chi^2}$ when this occurs by setting the effective mass $m_\chi$ to zero and solving to obtain
\begin{align}
    \expval{\chi_{\rm crit}^2}=\frac{2b}{3q}f\expval{\phi}\label{eq:chicrit}~.
\end{align}
When $b$ is close to one and $q \gg 1$, tachyonic resonance typically drives $\phi\to\chi\chi$ decays until around half the energy of the inflaton field is converted to $\chi$. At this point, the backreaction of $\chi$ on the inflaton field halts further energy transfer.

%%%%%%%%%%%%%%%%%%%%%%%%%%%%%%%%%%%%%%%%%
\subsection{Spillway Preheating}
\label{subsec:model2}
%%%%%%%%%%%%%%%%%%%%%%%%%%%%%%%%%%%%%%%%%
The spillway preheating model in~\cite{Fan:2021otj} aims to improve the energy depletion of the inflaton in the tachyonic resonance model by coupling a fermion field $\psi$ to the scalar daughter particle. The full potential of the theory is now
\begin{align}
    \label{eq:spillwaypotential}
    V (\phi, \chi, \psi)=\frac12 m^2\phi^2 +\frac12 \frac{M^2}f\phi\chi^2 +\frac14\lambda\chi^4+y\chi \bar\psi\psi~.
\end{align}
The added Yukawa interaction allows for perturbative decays of $\chi\to\bar\psi\psi$. The purpose of this addition is to deplete the $\chi$ particles, which consequently reduces the backreaction of $\chi$ against the inflaton condensate. This allows for more $\phi\to\chi\chi$ decays which will improve the energy transfer from the inflaton to radiation. At tree-level, the decay width of $\chi\to\bar\psi\psi$ is 
\begin{align}
    \label{eq:spillwaydecaywidth}
    \Gamma_{\chi}=\frac{ y^2 m_\chi}{8\pi}~,
\end{align}
where $m_\chi$ is the effective mass of $\chi$, defined as the curvature about the minimum of its potential and given by
\begin{align}
    \label{eq:chimass}
    m_\chi(\phi)=\begin{cases}
        \sqrt{\frac{M^2}{f}\phi}&\phi>0\\
        \sqrt{\frac{2M^2}{f}|\phi|}&\phi<0~.\\
    \end{cases}
\end{align}

\begin{figure}[h]
    \centering
    \includegraphics[width=7.3cm]{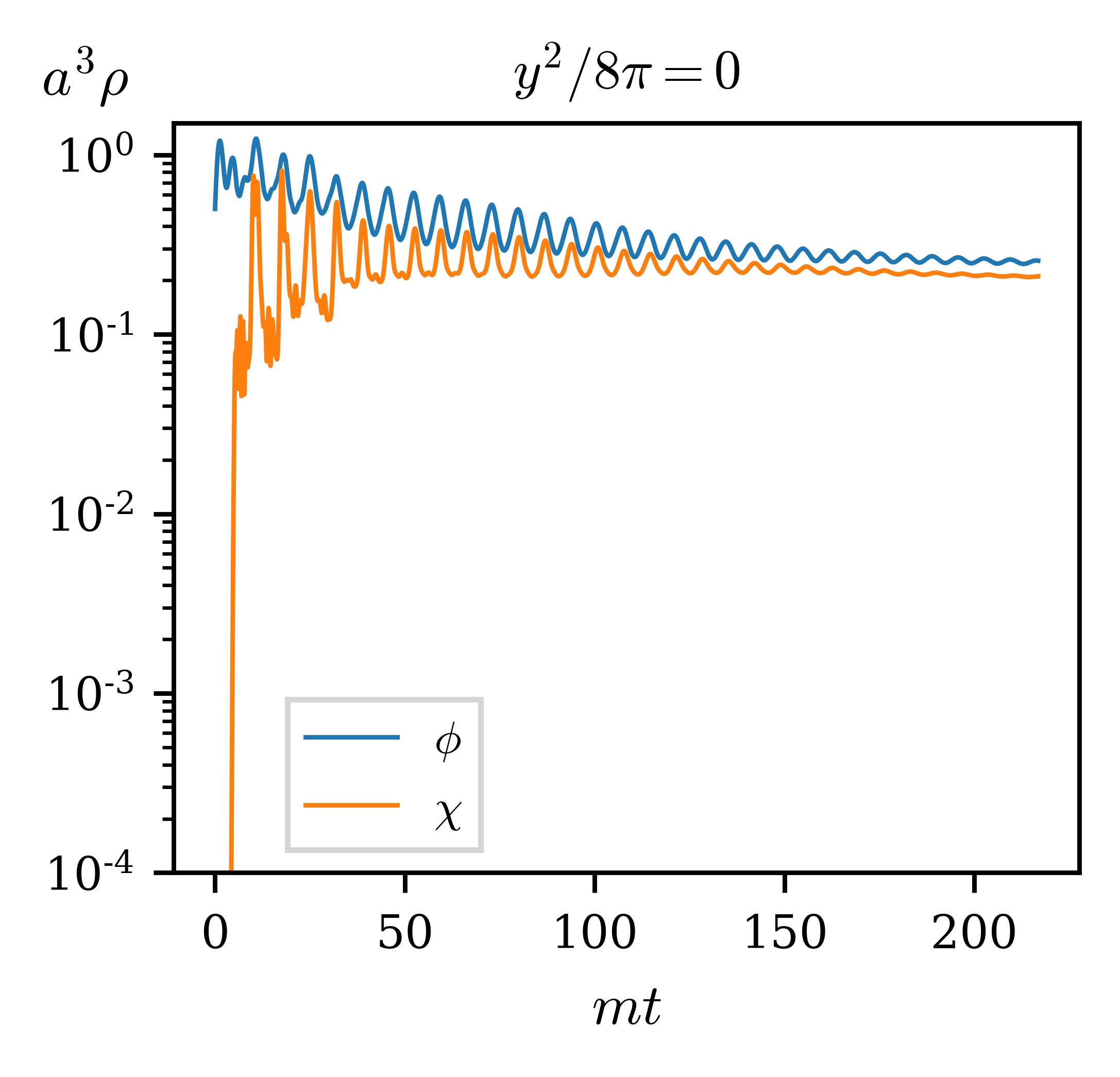}
    \includegraphics[width=7.3cm]{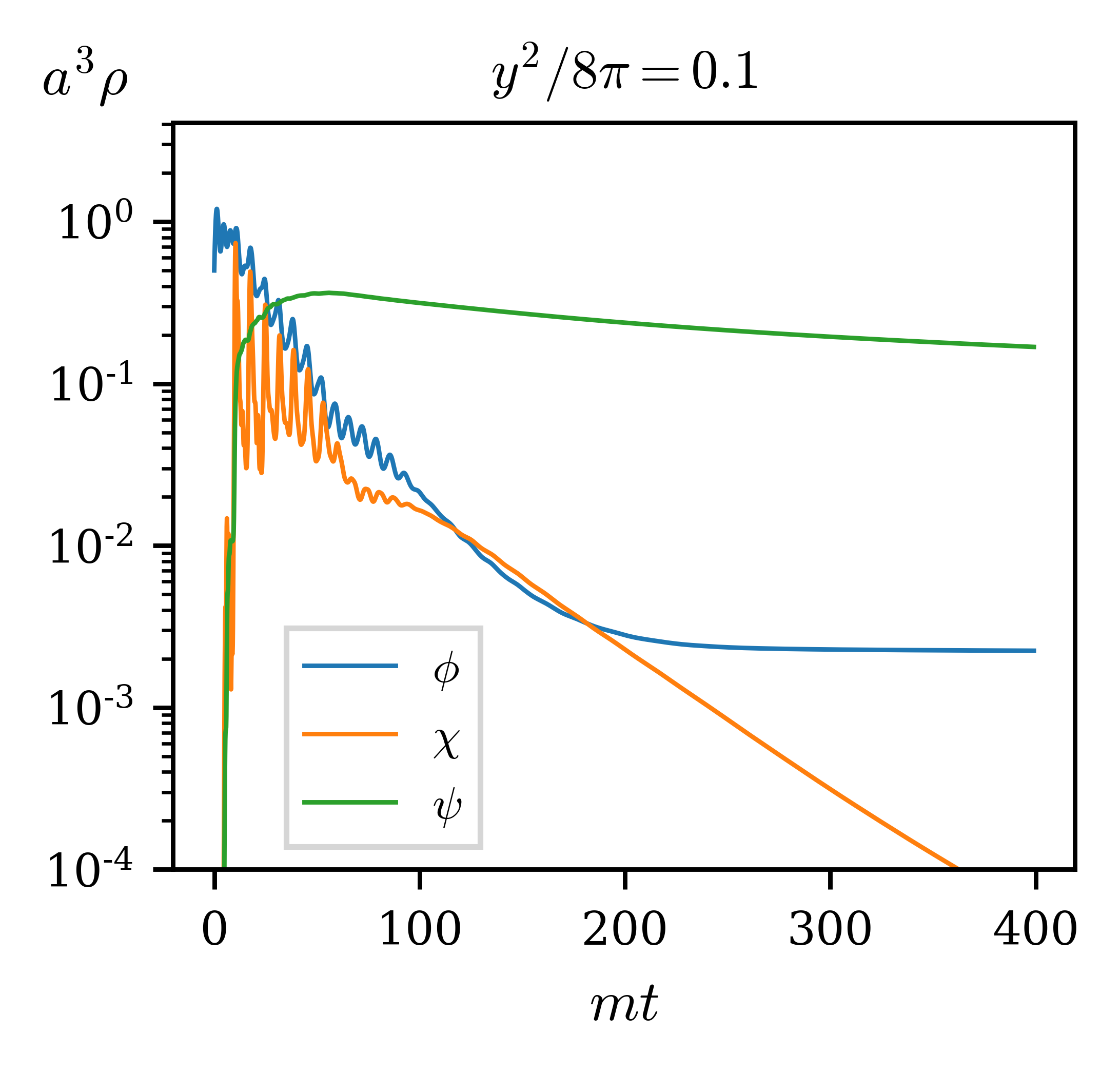}
    \caption{Comparison of the comoving energy densities for different fields as a function of time in the tachyonic resonance (left) and spillway preheating (right) models. We set $q=100$, $b=0.5$ for both panels. For the spillway preheating, we fix $y^2/(8\pi)=0.1$.}
    \label{fig:energy}
\end{figure}

As an example, we show in \Fig{fig:energy} one benchmark numerical result of the comoving energy densities of the fields evolving over time in the tachyonic resonance and spillway preheating models. It demonstrates an improved energy transfer of the spillway model by about two orders of magnitude compared to that of the tachyonic resonance model. $\chi$ and $\psi$ are both radiation-like and their energy densities scale as $a^{-4}$ with $a$ the scale factor, while the inflaton energy density scales as $a^{-3}$. As a result, the evolution of the equation of state of the whole system could be quite different in both models, with consequences which we will discuss more in the following sections.   

\break

%%%%%%%%%%%%%%%%%%%%%%%%%%%%%%%%%%%%%%%%%
\section{Observable 1: Inflationary Observables}
\label{sec:obs1}
%%%%%%%%%%%%%%%%%%%%%%%%%%%%%%%%%%%%%%%%%
In this section, we will investigate the effects of spillway preheating on two inflationary observables: the scalar spectrum index $n_s$ and tensor-to-scalar ratio $r$. We will first briefly review the basic formalism of these two observables, which are sensitive to the average equation of state $\bar{w}_{\text{re}}$ before reheating completes. Then we will present the dependence of $\bar{w}_{\text{re}}$ on the key parameters in the spillway preheating scenario. Lastly, we will show the allowed values of $n_s$ and $r$ in the spillway preheating model, comparing them with the current and near-future measurements.

%%%%%%%%%%%%%%%%%%%%%%%%%%%%%%%%%%%%%%%%%
\subsection{$n_s$ and $r$}
\label{subsec:nsr}
The power spectrum for scalar perturbations generated during inflation, $\Delta^2_{\mathcal R}(k)$, is given by
\begin{align}
    \Delta^2_{\mathcal R}(k)=\left(\frac{H^2}{2\pi\dot\phi }\right)^2\Big|_{k=aH}=\frac{1}{8\pi^2\epsilon_V}\frac{H^2}{M_{\text{pl}}^2}~,
\end{align}
where $H$ is the Hubble scale when the mode with wave number $k$ exits the horizon. The second equality is obtained by substituting in the potential slow-roll parameter $\epsilon_V = \frac{M_{\rm pl}^2}{2} (V(\phi)^\prime/V(\phi))^2$ with $V(\phi)$ being the inflaton potential during inflation and $V^\prime$ the derivative with respect to $\phi$. We also use the attractor solution that the inflaton velocity during inflation is given by $\dot{\phi}=-V^\prime/(3H)$. This has a weak scale-dependence described by a power law
\begin{align}
    \Delta_{\mathcal R}^2(k)=A_s\left(\frac{k}{k_*}\right)^{n_s-1}~,
\end{align}
with $k_*$ a reference scale and $n_s$ the scalar spectrum index, which is related to the potential slow-roll parameters $\epsilon_V$ and $\eta_V$ as:
\begin{align}
    \label{eq:nsslowroll}
    n_s-1\equiv \dv{\log\Delta^2_{\mathcal R}}{\log k}=-6\epsilon_V (\phi_k)+2\eta_V(\phi_k)~,
\end{align} 
where $\eta_V = M_{\rm pl}^2 V^{\prime \prime}/V$ and $\phi_k$ denotes the inflaton field value at the $k$-horizon exit. The scalar amplitude $A_s$ could be expressed in terms of the slow-roll parameters as:
\begin{align}
    \label{eq:As}
    A_s=\frac{1}{24 \pi^2} \frac{V}{M_{\rm pl}^4 \epsilon_V}~.
\end{align}
In summary, the scalar perturbation spectrum is completely determined by the slow-roll parameters $\epsilon_V$ and $\eta_V$.

The tensor perturbation of the metric, on the other hand, has the power spectrum
\begin{align}
    \Delta_h^2(k)=\frac{2}{\pi^2}\frac{H^2}{M_{\text{pl}}^2}\Big|_{k=aH}~.
\end{align}
Then the tensor-to-scalar ratio $r$ is determined by
\begin{align}
    \label{eq:rslowroll}
    r=\frac{\Delta_h^2(k)}{\Delta_{\mathcal R}^2(k)} =16\epsilon_V(\phi_k)~.
\end{align}

To predict the observables $n_s$ and $r$, it suffices to determine the slow-roll parameters $\epsilon_V(\phi_k)$ and $\eta_V(\phi_k)$ at $k$-horizon exit. To achieve that, we need to specify an inflationary model with a chosen $V(\phi)$. Then the only remaining parameter to determine is $\phi_k$. $\phi_k$ is related to $N_k$, the number of $e$-folds from the $k$-horizon exit to the end of inflation, by the equation 
\begin{align}
    \label{eq:Nkslowroll}
    N_k&=\int_{\phi_k}^{\phi_e}\frac{|\dd\phi|}{\sqrt{2\epsilon_V}}~.
\end{align}
Here $\phi_e$ is the value of the inflaton at the end of inflation, and $\epsilon_V(\phi_e)$ is the potential slow-roll parameter when inflation ends, which we take to be one. 
$N_k$ is also related to the $e$-folds between the end of inflation and today in a given expansion history and could be computed as~\cite{Liddle:2003as}
\begin{align*}
N_k=&-\log\frac{k}{a_0T_0}-\frac{1-3\bar{w}_{\rm re}}4N_{\rm re}-\frac14\log\frac{30}{g_{\rm re}\pi^2}-\frac13 \log\frac{11g_{s,\rm re}}{43}\\&+\frac14\log\frac{\pi^2 r A_s}{6}+\frac14\log\frac{V(\phi_k)}{V(\phi_e)}+\frac14\log\frac23~,\numberthis \label{eq:Nkconstraint}
\end{align*}
where $a_0=1$ and $T_0=2.73$ K are the present-day scale factor and CMB temperature respectively. We follow the Planck 2018 convention and use a reference scale of $k=0.05\,\text{Mpc}^{-1}$. $g_{\rm re}\approx 10^2$ is the effective degree of freedom for energy density while $g_{s, \rm re}\approx 10^2$ is that for entropy during reheating. We use the Planck 2018 measurement value of  $A_s=2.1\times 10^{-9}$ \cite{{Akrami:2018odb}}. Two important parameters characterizing reheating enter this equation: $N_{\rm re}$ is the number of $e$-folds between the end of inflation and the end of reheating when the universe becomes radiation-dominated, and $\bar{w}_{\rm re}$ is the average equation of state of the universe during the entire reheating period,
\beq
\bar{w}_{\rm re} = \frac{1}{N_{\rm re}}\int_{N_{\rm re}} w(N')\dd N'~.
\eeq
\Eq{eq:Nkconstraint} thus gives us a relation between the observables we would like to determine, $r$ and $n_s$, and the ones we could directly compute from a given reheating model, $\bar{w}_{\rm re}$ and $N_{\rm re}$.

The cosmological equation of state at any given point of time is $w_{\rm re}=p_{\rm tot}/\rho_{\rm tot}$, where $p_{\rm tot}$ is the average pressure of the entire system and $\rho_{\rm tot}$ is the corresponding space-averaged comoving energy density. 
The value of $w_{\rm re}$ clearly depends on the preheating dynamics. The inflaton field is matter-like, corresponding to an equation of state $w=0$. The daughter fields in the spillway model, $\chi$ and $\psi$, are radiation-like, corresponding to $w=1/3$. Because the distribution of energy densities for different species varies across preheating models, we expect different preheating models to generate different time evolutions of $w_{\rm re}$ between 0 and $1/3$, before the universe completes reheating and reaches a constant $w_{\rm re} = 1/3$.

The number of $e$-folds between the end of inflation and the end of reheating, $N_{\rm re}$, could be calculated as 
\begin{align}
  N_{\rm re}=\frac{1}{3(1+\bar{w}_{\rm re})}\log(\frac{\rho_{0}}{\rho_{\rm re}})~,  
\end{align}
where $\rho_0=m^2 f^2/2$ is the inflaton energy density at the end of inflation. $\rho_{\rm re}$ is the energy density at the end of reheating when the perturbative decays, $\phi \to \chi \chi$, completes the conversion from the remaining inflaton energy density to radiation~\cite{Kofman:1997yn}:
\begin{align}
 \rho_{\rm re}=3\Gamma_{\phi}^2M_{\text{pl}}^2~,\quad   \Gamma_{\phi}=\frac{M^4}{8\pi m f^2}~.
\end{align}
Thus once we know $\bar{w}_{\rm re}$, $N_{\rm re}$ is completely determined. Then combining \Eq{eq:nsslowroll}, \Eq{eq:rslowroll}, \Eq{eq:Nkslowroll}, and \Eq{eq:Nkconstraint}, we could determine $r$ and $n_s$ for a given inflaton potential. 
In the next section, we will discuss how to compute $\bar{w}_{\rm re}$.

%%%%%%%%%%%%%%%%%%%%%%%%%%%%%%%%%%%%%%%%%
\subsection{Computations of $\bar{w}_{\rm re}$}
\label{subsec:model1sim}
%%%%%%%%%%%%%%%%%%%%%%%%%%%%%%%%%%%%%%%%%
 As described in the last section, it is the average equation of state $\bar{w}_{\rm re}$ that enters the computation of inflationary observables, $n_s$ and $r$. However, to make the physical effect of the spillway mechanism transparent, we will first discuss the effect of the spillway on the entire evolution of $w_{\rm re}$ between the end of inflation and the end of reheating, and then present the resulting variation in $\bar{w}_{\rm re}$. 

A combination of numerical simulations and analytical estimates will be necessary to understand the effect of the spillway on $\bar{w}_{\rm re}$: numerical simulations are needed to capture the nonperturbative dynamics shortly after inflation, while analytical estimates will be needed to extrapolate what we learn from the numerical simulations until the end of reheating, which is too long to be simulated completely.

\subsubsection{Numerical Simulations}

We follow the same numerical approach as in~\cite{Fan:2021otj} to simulate the time evolution of the system in the spillway scenario. The equation of motion for the inflaton is
\begin{align}
    \label{eq:phieom}
    \ddot\phi + 3H\dot\phi -\frac1{a^2}\nabla^2 \phi+m^2\phi+\frac12\frac{M^2}{f}\chi^2=0~,
\end{align}
while its direct scalar daughter is governed by 
\begin{align}
    \label{eq:chieom}
    \ddot\chi + 3H\dot\chi -\frac1{a^2}\nabla^2 \chi+\lambda\chi^3+\frac{M^2}{f}\phi\chi+\Gamma_\chi\dot\chi=0~,
\end{align}
where the perturbative decays of $\chi$ to the fermions serve as the friction term. The fermionic decay product is approximated as a perfect, homogeneous, radiation-like fluid whose energy density is governed by
\begin{align}
    \label{eq:psieom}
    \dot\rho_\psi+4H\rho_\psi-\expval{\Gamma_\chi\dot\chi^2}=0~,
\end{align}
where $\Gamma_\chi\dot\chi^2$ acts as a source term. This system of equations, together with the Friedmann equations to compute the evolution of the scale factor, could be solved using the LatticeEasy software package \cite{Felder:2000hq} with the integrator replaced with the fourth-order Runge-Kutta algorithm. The fields are simulated on a $128^3$ lattice of width 2$m^{-1}$. For each simulation we fix the initial oscillation amplitude $f=M_{\text{pl}}$ and the inflaton mass $m=10^{-6}M_{\text{pl}}$. We note that of the two Friedmann equations,
\begin{align}
    \frac{\ddot{a}}{a}&=-\frac{4 \pi G}{3}\left\langle\rho_{\text {tot}}+3 p_{\text {tot}}\right\rangle\label{eq:friedmann1}~,\\
    \left(\frac{\dot{a}}{a}\right)^2&=\frac{8 \pi G}{3}\left\langle\rho_{\text {tot}}\right\rangle~,\label{eq:friedmann2}
\end{align}
only one equation is necessary to evolve the scale factor. $G$ above is the Newton constant. LatticeEasy chooses to use \eqref{eq:friedmann1}, and \eqref{eq:friedmann2} is computed as a consistency check for the conservation of energy.

%\hspace{0pt}

In the scenario we are interested in, preheating through non-perturbative particle production is effective before the perturbative reheating completes the transition from inflaton domination to thermal big bang. Thus we  require that the inflaton's perturbative decays happen (much) later than the timescale of preheating. For efficient preheating, it happens almost immediately after inflation so we require $\Gamma^{-1}_{\phi} \gg 100m^{-1}$.  This roughly corresponds to $q\leq 2000$. 
Additionally, we do not test values of $q$ smaller than 10 or values of $b$ smaller than $0.1$, as under these conditions preheating is always inefficient before reheating kicks in. 
Lastly we only consider $y^2/8\pi\leq 0.1$: further increasing from $y^2/8\pi=0.1$ to around $y^2/8\pi=0.7$ results in little change to the simulation results, and the coupling enters a regime where the perturbative computation is no longer valid. In summary, we run simulations for values of $10\leq q\leq 2000$, $0.1\leq b<1$, and $0\leq y^2/8\pi\leq 0.1$.

%%%%%%%%%%%%%%%%%%%%%%%%%%%%%%%%%%%%%%%%%
\subsubsection{Evolution of $w_{\rm re}$}
%%%%%%%%%%%%%%%%%%%%%%%%%%%%%%%%%%%%%%%%%

In this section, we analyze the effect of the spillway mechanism (turning on $y\neq 0$) on the evolution of the equation of state of the universe, based on the simulation method described previously. The evolution of the equation of state divides into three scenarios depending on the value of $q$.

\begin{figure}[h]
    \centering
    \includegraphics[width=7cm]{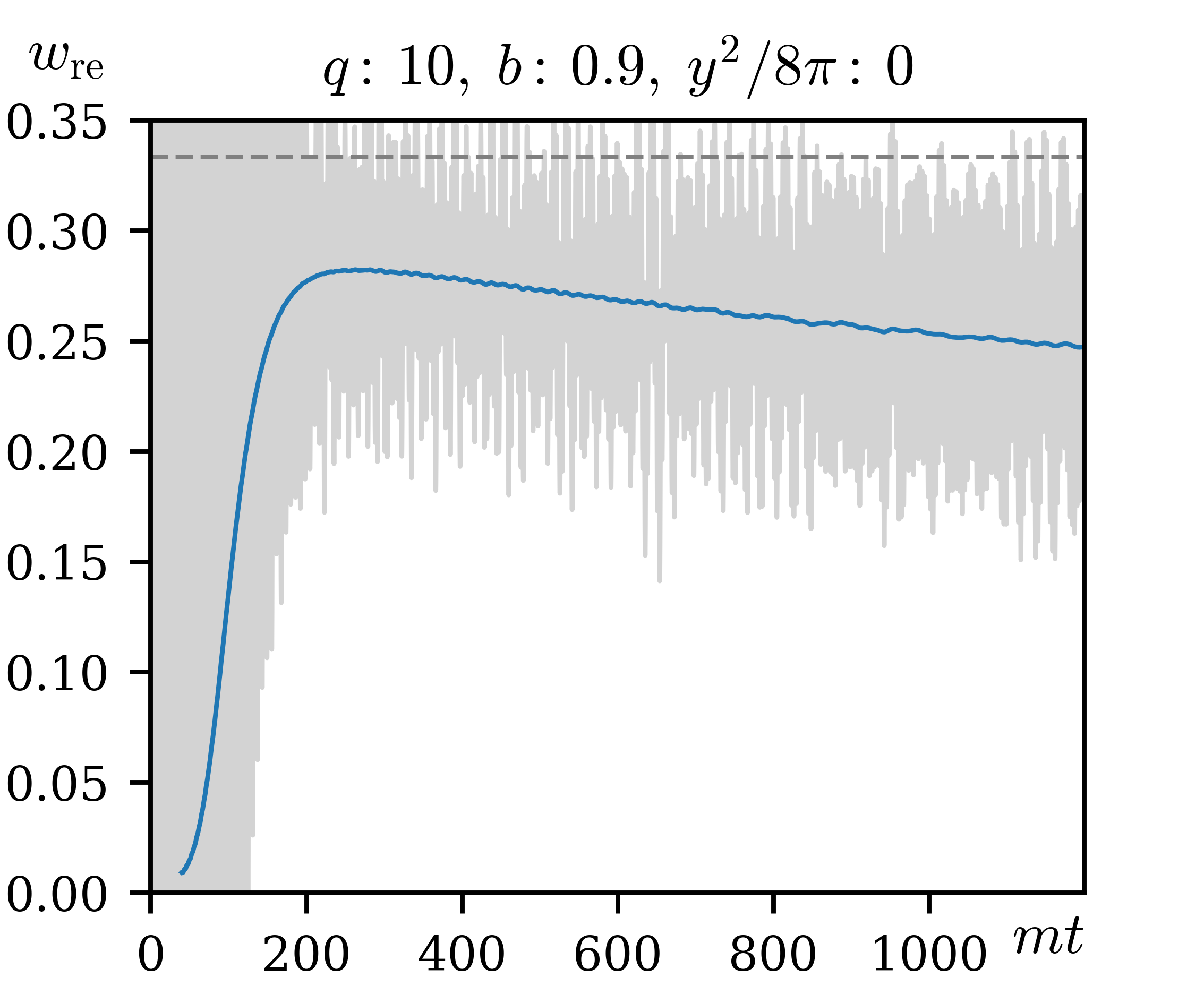}
    \includegraphics[width=7cm]{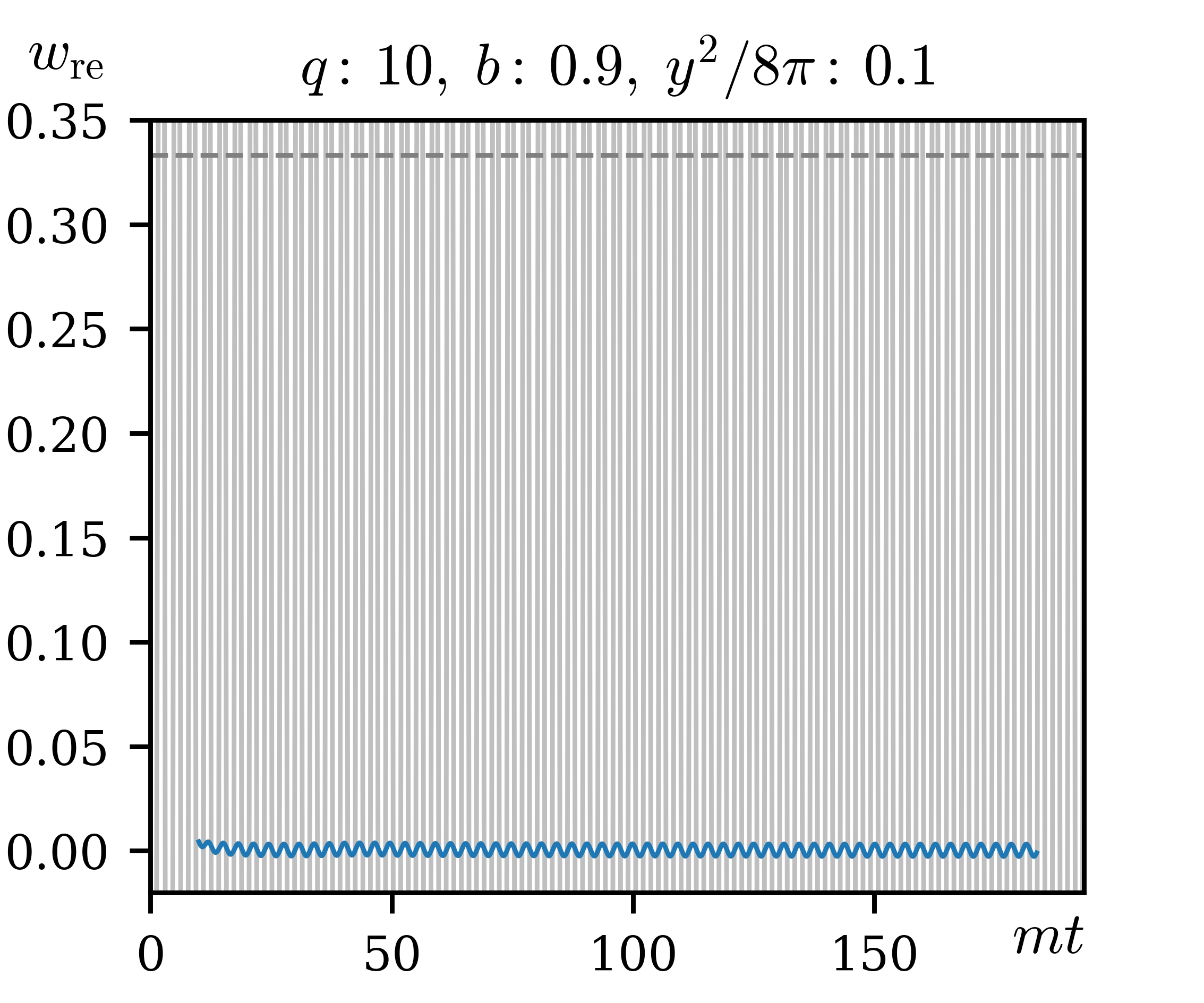}
 \caption{$w_{\rm re}$ (gray) and a time-averaged $w_{\rm re}$ (blue) as functions of time $mt$, for $y=0$ (left) and $y^2/(8\pi) =0.1$ (right) at $q=10$ and $b=0.9$. The dotted horizontal gray line indicates $w_{\rm re}=1/3$.}\label{fig:inefficient}
\end{figure}

For low values of $q\sim 10$, turning on the spillway with a large value of $y$ prevents efficient particle production all together, because the $\chi$ production from preheating is too slow compared to the $\chi\rightarrow \bar{\psi}\psi$ decay, and resonant production of $\chi$ is shut off prematurely. In contrast, for tachyonic resonance with $y = 0$, resonant production of $\chi$ happens without being hindered. The difference in the resonant production of $\chi$ is directly reflected in the evolution of equation of state in the two models. \Fig{fig:inefficient} shows the time evolution of $w_{\rm re}$ for $b=0.9$ and $q=10$. We find that the tachyonic resonance model with no spillway shows an initial increase of $w_{\rm re}$ to around $0.27$, followed by a gradual decrease. This implies that although the tachyonic resonance production functions for the first $\mathcal{O}(100)$ oscillations, the backreaction from the produced $\chi$ particles will eventually slow down the production. Then the $\chi$ production is not rapid enough to keep up with the redshifting of radiation due to the expansion of the universe, causing the system to relax back to a matter domination state.
In contrast, $w_{\rm re}$ remains fixed at zero in the spillway model at this low $q$ and the system never leaves the matter domination state, due to the lack of effective $\chi$ production.

\begin{figure}[h]
    \centering
      \includegraphics[width=7.3cm]{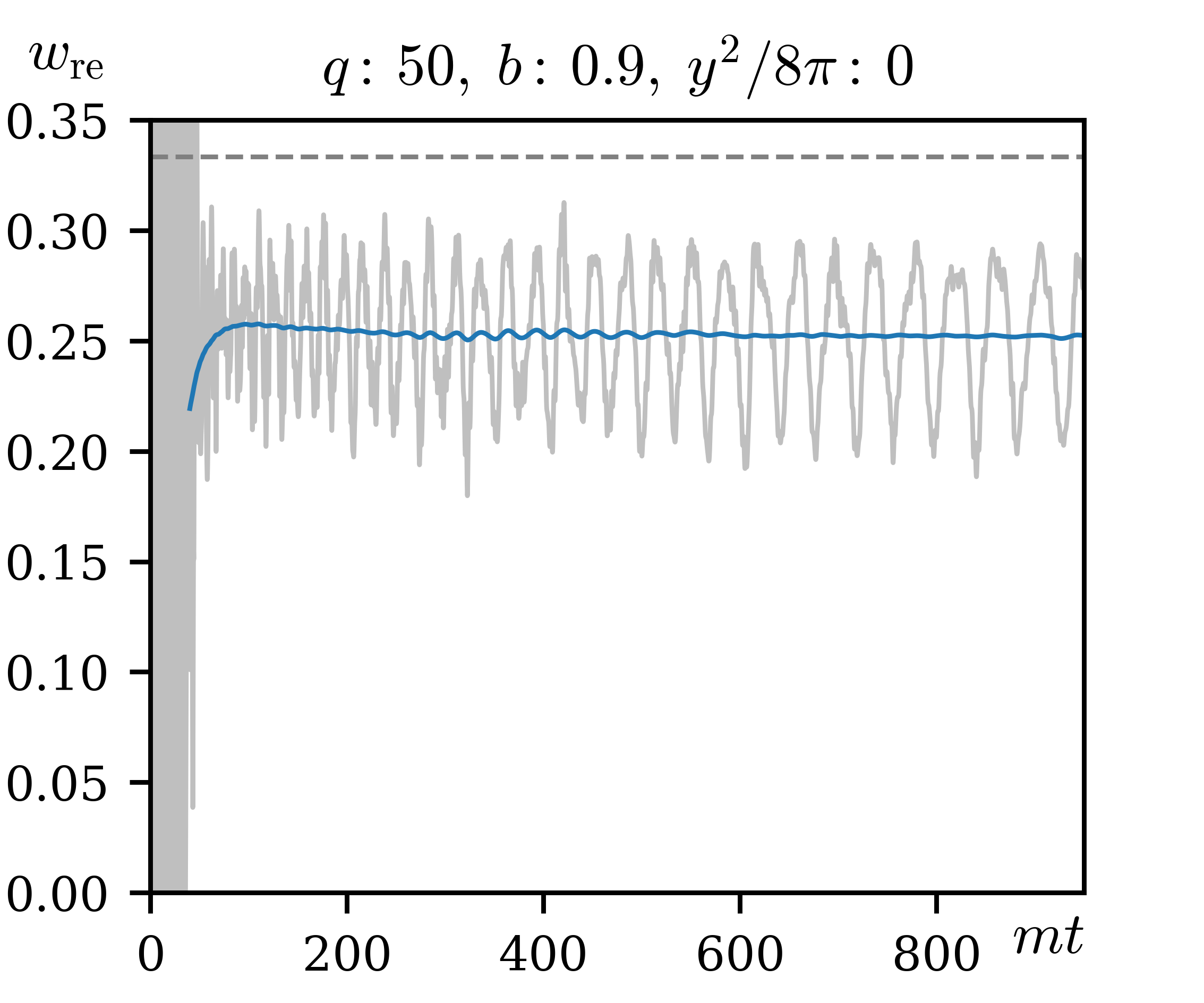}
          \includegraphics[width=7.3cm]{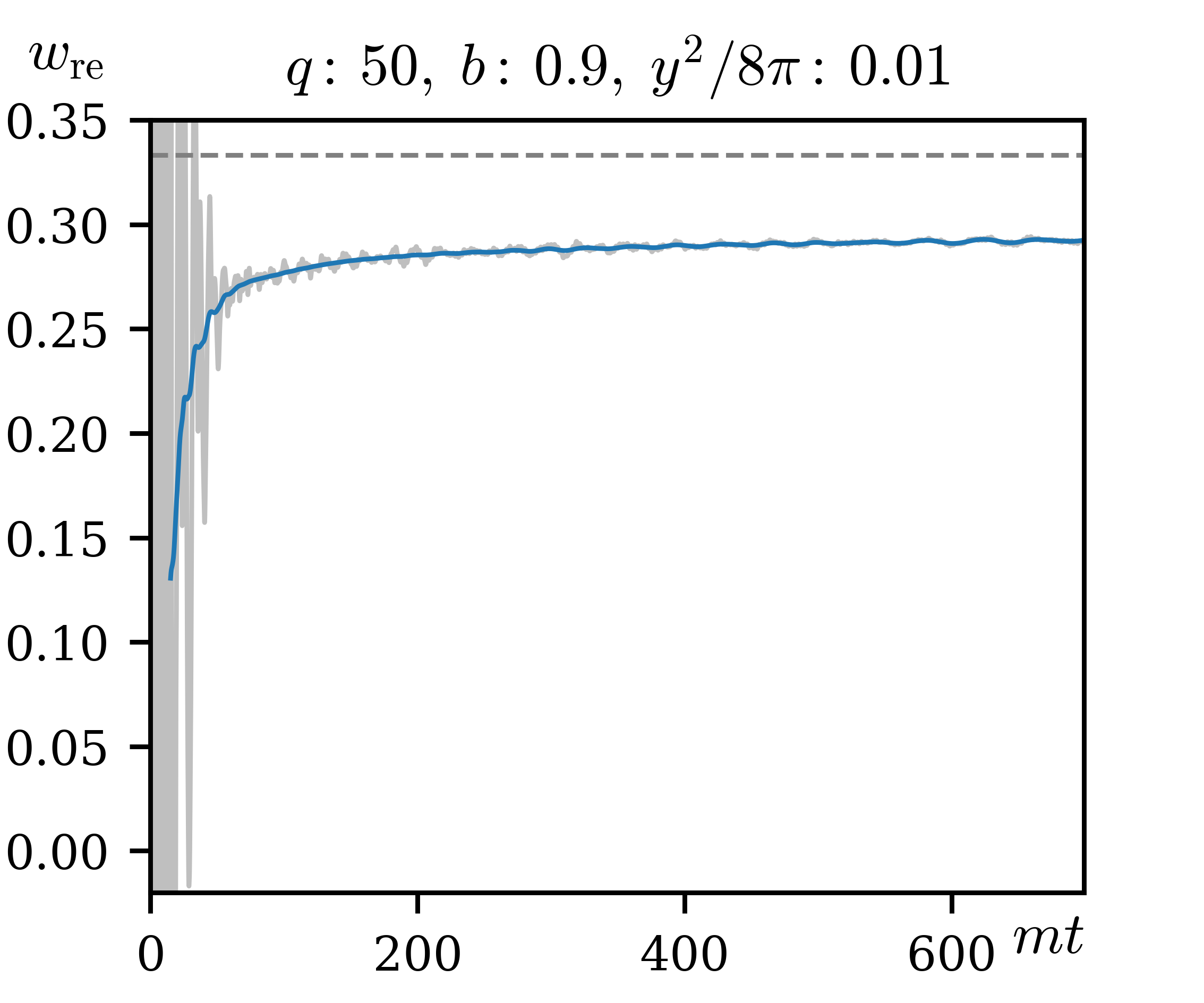}
    \includegraphics[width=7.3cm]{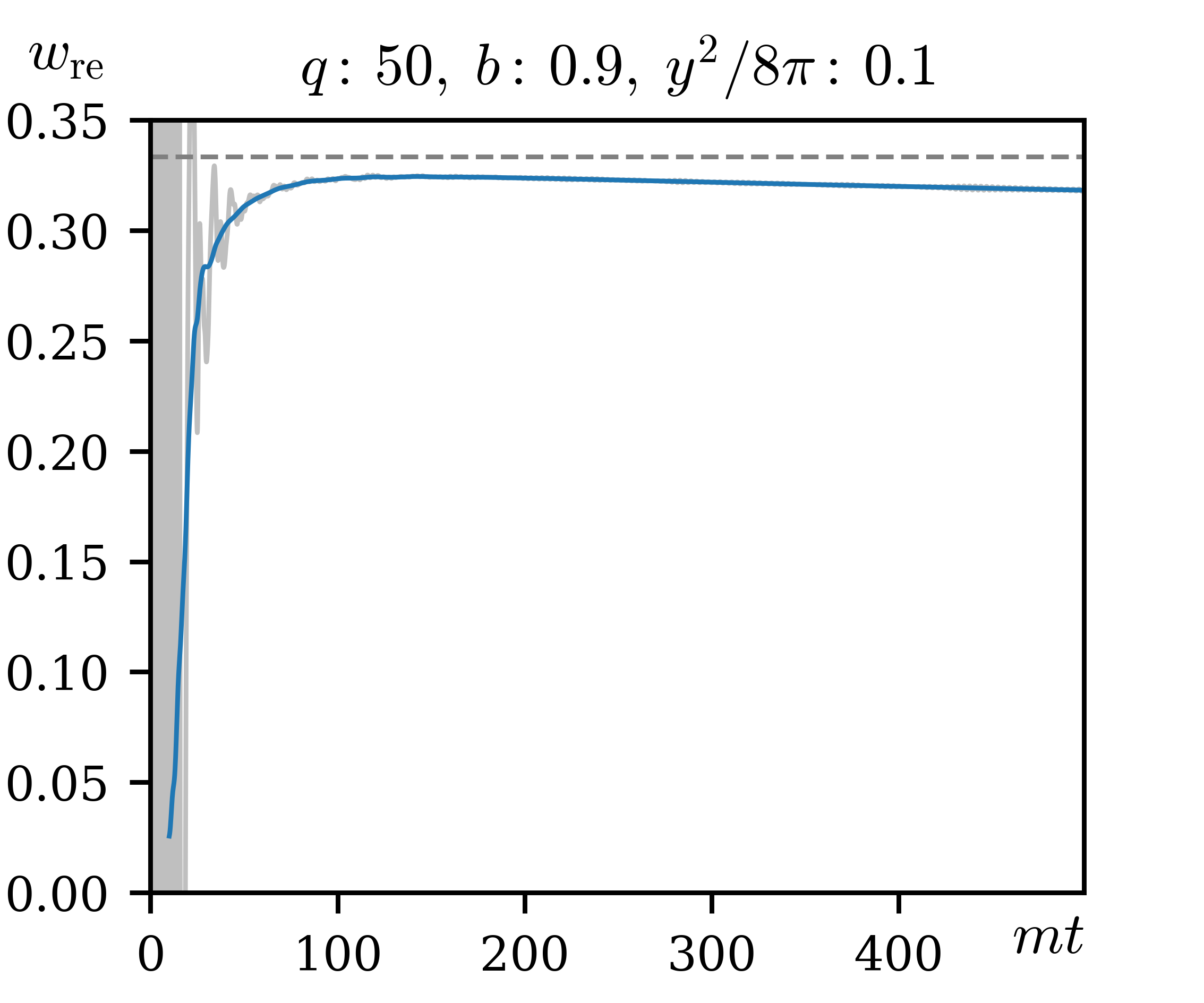}
\caption{$w_{\rm re}$ (gray) and a time-averaged $w_{\rm re}$ (blue) as functions of time for $q=50$. The dotted gray line indicates $w=1/3$. We fix $b=0.9$ and choose different $y$'s for the three panels.}\label{fig:efficient}
\end{figure}

For the medium range of $q$ ($10\lesssim q \lesssim 1000$), turning on spillway with greater values of $y$ has two main effects: the first is increasing the maximum value of equation of state achieved to be closer to radiation-like, $w_{\rm re} \approx 1/3$. Spillway is known to make energy transfer from the inflaton to its daughter particles more efficient when $q$ is sufficiently large. Since the daughter particles are radiation-like, the equation of state will be closer to $1/3$, as expected. In \Fig{fig:efficient}, we present the time evolution of $w_{\rm re}$ for $q=50$, $b=0.9$ and three different $y$'s. All three benchmarks show similar behavior wherein $w_{\rm re}$ increases to a particular value and remains roughly constant for $\mathcal O(100)$ oscillations, then gradually declines. The plateau-like behavior reflects that the larger $q$ is, the more efficient $\phi\to\chi\chi$ production is, and $\chi$ is frequently replenished to counteract the decrease of radiation-like energy density due to redshift. The peak value of $w_{\rm re}$ becomes more radiation-like as we increase the $y$ coupling, reaching an almost completely radiation-like equation of state at $y^2/(8\pi)=0.1$. In the tachyonic case with $y=0$, the system reaches a mixed matter-radiation state around $w_{\rm re}\approx 0.26$. 

\begin{figure}[h]
    \centering
    \includegraphics[width=7cm]{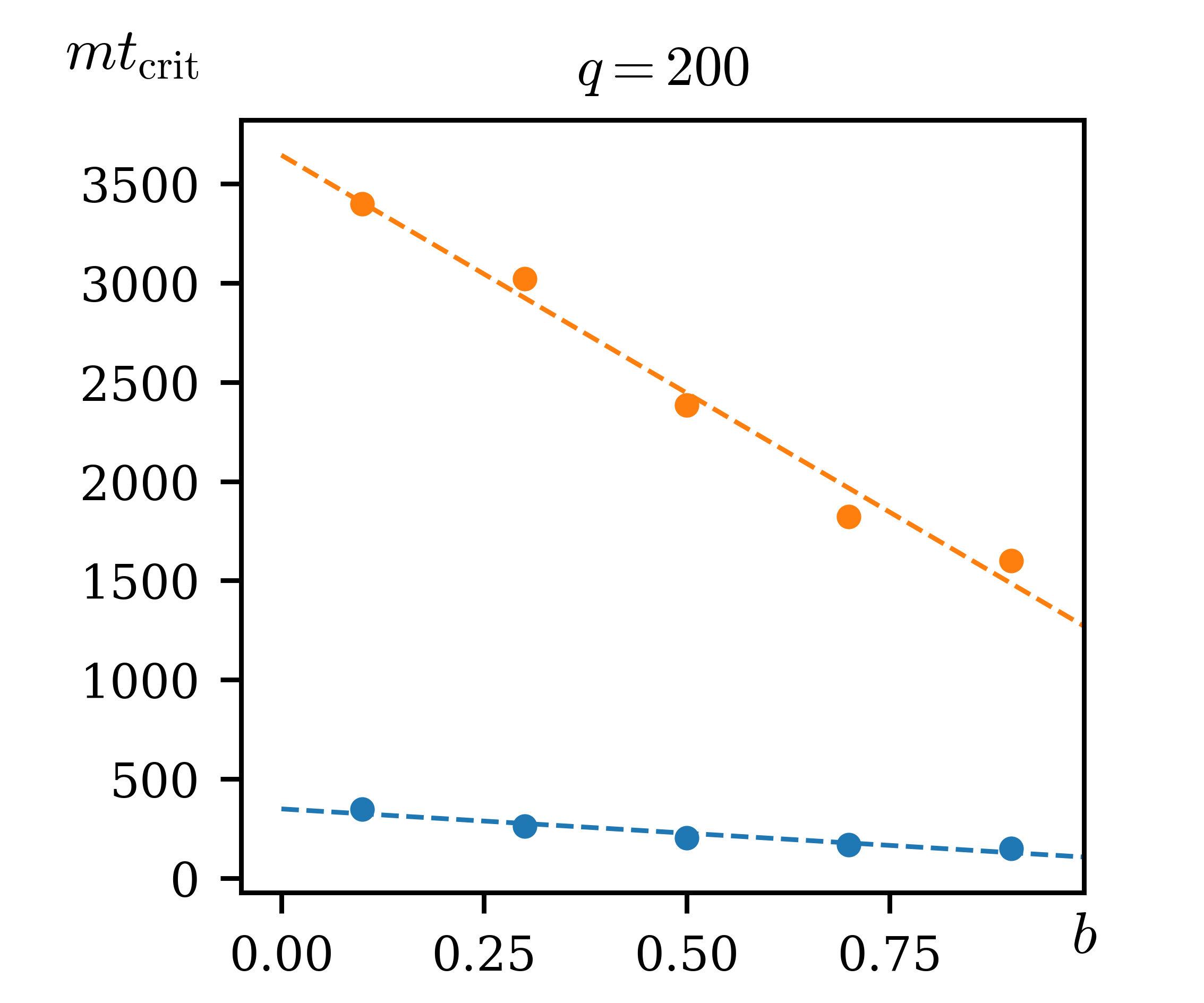}
    \includegraphics[width=7cm]{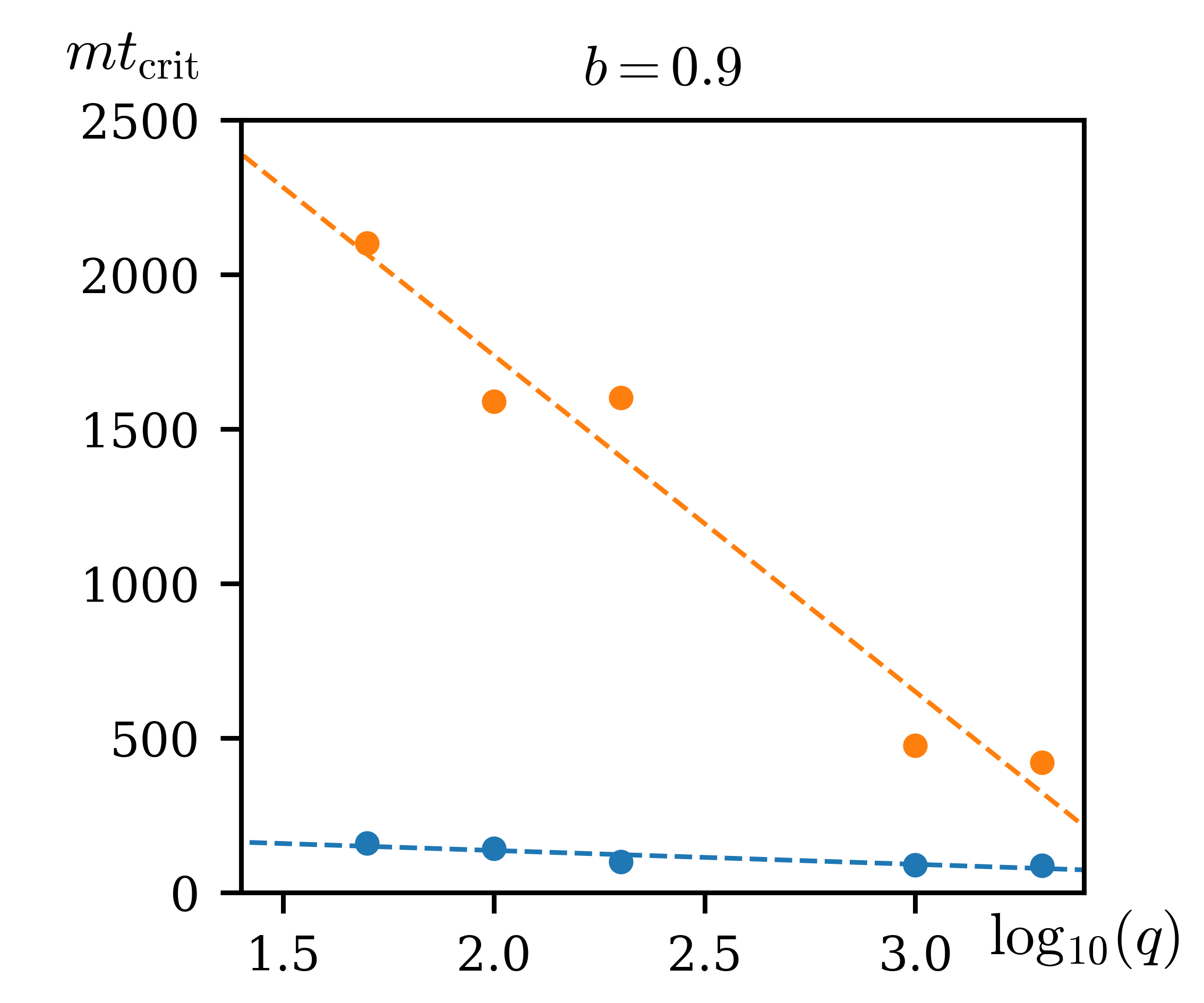}
    \caption{Critical time at which redshift domination occurs as a function of $b$ (left) and $q$ (right). Orange denotes $y^2/8\pi=0.01$, while blue denotes $y^2/8\pi=0.1$. Dots indicate simulation results, and dashed lines are lines of best fit. When depletion of the inflaton energy density is faster (corresponding to larger $q$, $b$, and $y$), the inflaton and its daughter particle stops interacting earlier, leading to a lower $t_{\rm crit}$.}
    \label{fig:tc}
\end{figure}

The second effect of turning on the spillway is related to the time evolution of the equation of state. When the value of $y$ is larger, the equation of state will decrease towards 0 from its maximum value \textit{earlier}.  In Fig. \ref{fig:efficient}, we notice a mild gradual decreasing trend in the equation of state in the $y^2/8\pi=0.1$ case, beginning at around $mt\approx 150$. Around this time, $\phi$ and $\chi$ have mostly been depleted, and the majority of the system's energy lies within $\psi$. $\psi$ has no backreaction on $\chi$ or $\phi$, so from this point onward, the time evolution of $w_{\text{re}}$ is dominated by redshift, which will slowly bring the system back to a matter-like equation of state. This trend is also present in the $y^2/8\pi=0.01$ case, but beginning much later around $mt\sim {\cal O}(1000)$. This is because when $y$ is smaller, the depletion of $\phi$'s and $\chi$'s energy densities is slower. The system remains interacting until later times, maintaining a constant equation of state. We denote by $t_{\rm crit}$ the time at which redshift becomes the dominant effect in the time evolution of $w_{\rm re}.$ In the tachyonic case, because depletion of the inflaton energy density is inefficient, the inflaton and scalar daughter field remains interacting over a much longer period, which is reflected by the oscillatory behavior of $w_{\rm re}$ in the $y=0$ case of Fig. \ref{fig:efficient}. While redshift pulls the system toward a matter-like equation of state, it never becomes the dominant effect, since any energy lost in $\chi$ is quickly replaced by $\phi\to\chi\chi$ decays. \Fig{fig:tc} plots the approximate dependence of the time $t_{\rm crit}$ against $y$, $b$, and $q$. We see a decrease in $t_{\rm crit}$ at larger values of $y^2/8\pi$, as discussed above. $t_{\rm crit}$ also decreases as both $b$ and $q$ increase, reflecting that increasing these parameters enhances the inflaton field decay rate, depleting the inflaton more quickly, and making the system enter redshift domination earlier. 

\begin{figure}[h]
    \centering
    \includegraphics[width=7cm]{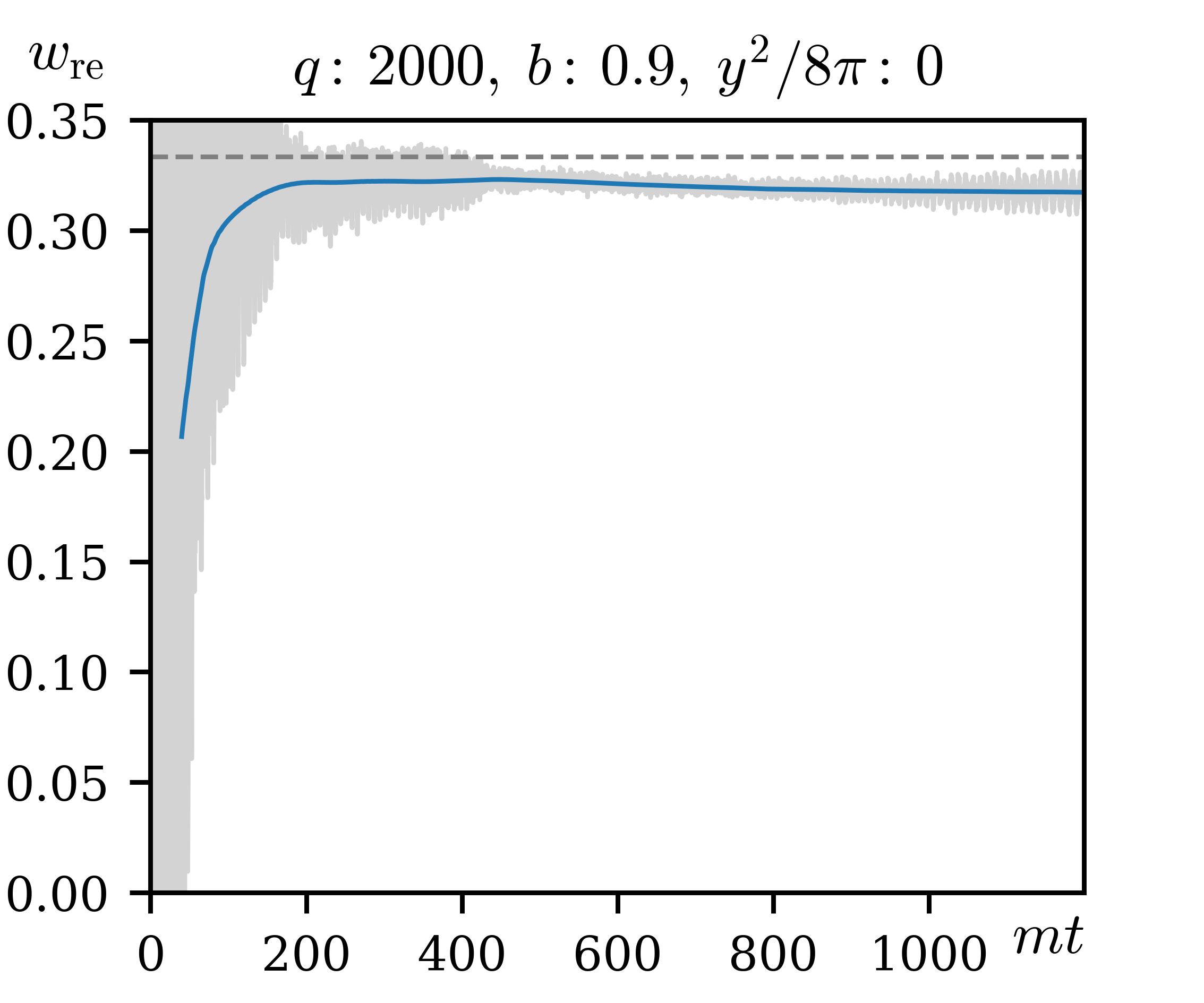}
    \includegraphics[width=7cm]{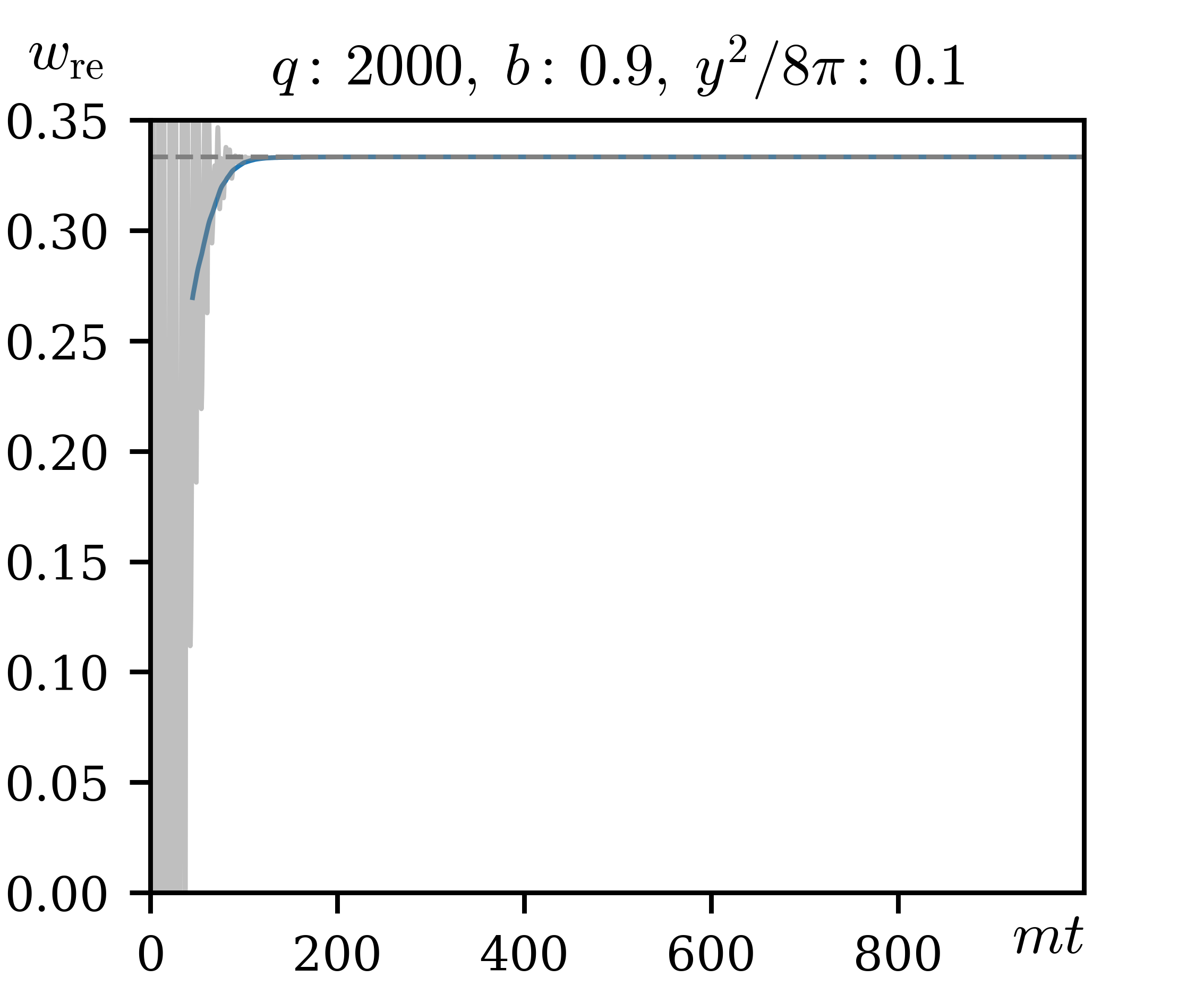}
    \caption{Equation of state and comoving energy densities for $q=2000$, $b=0.9$, $y^2/8\pi=0.1$.}
    \label{fig:q2000}
\end{figure}

At even higher values of $q \gtrsim 1000$, the spillway mechanism becomes so efficient that the equation of state stays radiation-like for a long time. \Fig{fig:q2000} compares $w_{\text{re}}$ for the tachyonic and spillway cases for $q = 2000$. The spillway model achieves a completely radiation-dominated equation of state, while the tachyonic potential reaches a slightly lower maximum of $w_{\rm re}\sim 0.32$. Beginning around $mt\sim 600$, the equation of state in the tachyonic case exhibits similar decreasing behavior observed at lower $q$ due to the redshifting of the radiation-like energy density. However, the spillway case remains at $w_{\rm re}=1/3$. This does not mean that the redshifting effect does not happen. In fact, $t_{\rm crit}$ is still $\mathcal{O}(100)m^{-1}$. But the equation of state stays very close to $w_{\rm re}=1/3$ even after $t_{\rm crit}$, because the inflaton field density around $t_{\rm crit}$ makes up only around a $\mathcal O(10^{-4})$ fraction of the total comoving energy density.

\begin{figure}[H]
    \centering
    \includegraphics[width=\textwidth]{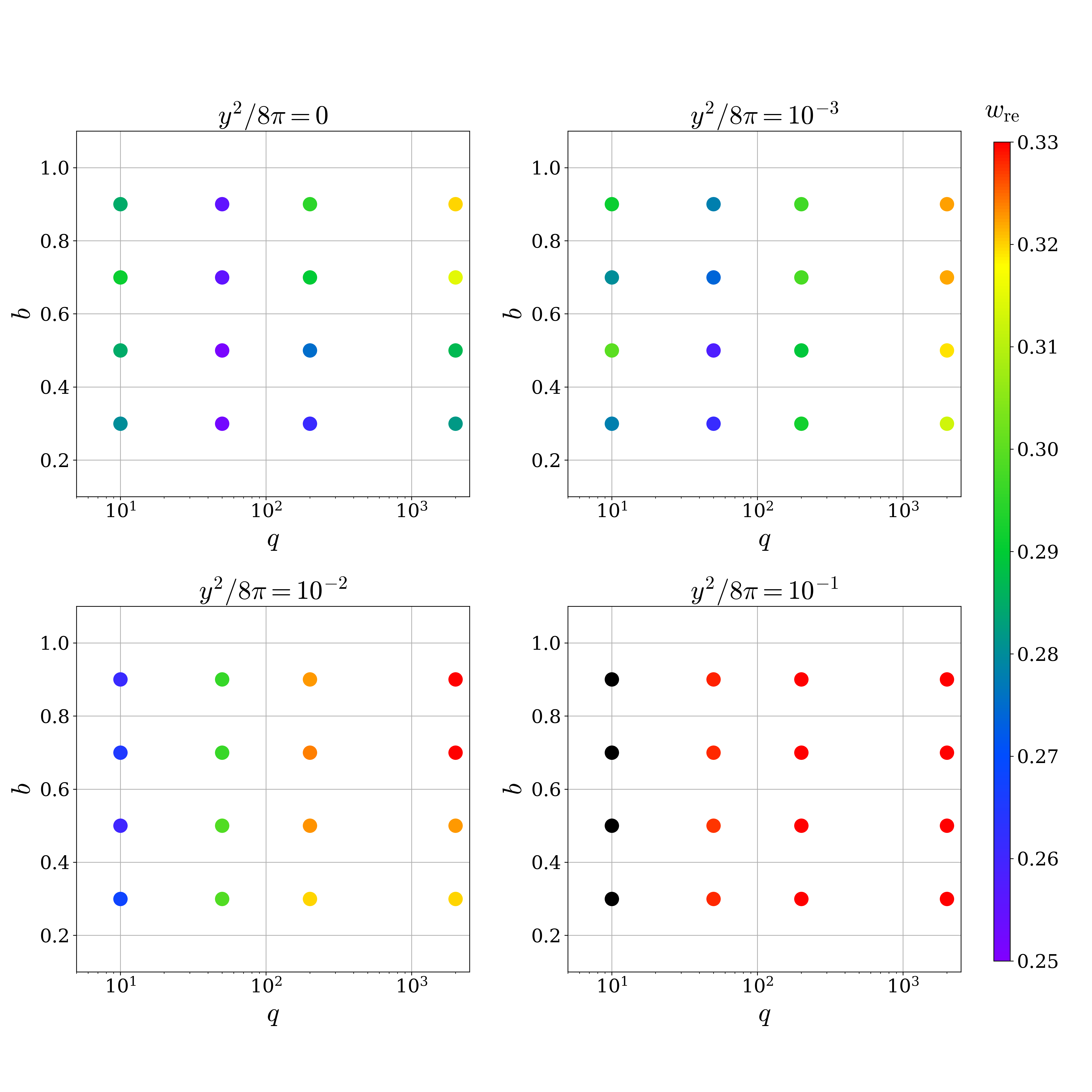}
    \caption{Peak value of $w_{\rm re}$ against $q$ and $b$ for different choices of $y$. The black dots in the panel of $y^2/(8\pi) = 0.1$ indicate no particle production at all.}
    \label{fig:wspace2}
\end{figure}

To gain a more comprehensive picture of the peak value of $w_{\rm re}$ as a function of the input parameters, we scan the three-dimensional parameter space  $y$, $q$ and $b$ and the results are presented in \Fig{fig:wspace2}. In \Fig{fig:wspace2}, we show peak values of $w_{\rm re}$ in the ($q, b$) plane, varying $y$. We see that in general, for a given $y$, the parameter space can be divided into two regimes. When $q>10$, $w_{\rm re}$ increases when $q$ and $b$ increases. This trend is most noticeable for $y^2/(8\pi) =10^{-2}$. This reflects that increasing the inflaton decay rate and decreasing the backreaction should both push the system to a more radiation-like state. For an even larger $y$, i.e., $y^2/(8\pi) = 0.1$,  the system reaches $w_{\rm re}\approx 1/3$ once $q>10$ and thus do not show any visible increase with increasing $q$ and $b$. We also observe that increasing $y$ above a threshold results in an increase in $w_{\rm re}$. For a small Yukawa coupling, e.g., $y^2/(8\pi) = 10^{-3}$, the spillway effect is limited and the evolution of $w_{\rm re}$ remains similar to the $y=0$ case.
When $q\sim 10$, there is no clear trend in the peak value of $w_{\rm re}$, except that $w_{\rm re}$ decreases as we increase {$y^2/(8\pi)$ above $10^{-3}$}, because the spillway mechanism kills particle production due to the fast perturbative decays.

%%%%%%%%%%%%%%%%%%%%%%%%%%%%%%%%%%%%%%%%%
\subsubsection{Effect on $\bar{w}_{\rm re}$}
%%%%%%%%%%%%%%%%%%%%%%%%%%%%%%%%%%%%%%%%%
The immense computational cost prevents numerical simulations from being run completely between the end of inflation until the end of reheating when $H = \Gamma_{\phi}$. Therefore some analytical estimate is needed to extrapolate the time evolution of $w_{\rm re}$ we have learned from numerical simulations to compute $\bar{w}_{\rm re}$ between the end of inflation and the end of reheating.

Recall, that the average equation of state is in general defined as
\beq
\bar{w} = \frac{1}{\Delta N}\int w(N')\dd N'~.\label{eq:wint_gen}
\eeq
Given the average of equation of state during two time periods, $\bar{w}_1$ and $\bar{w}_2$, each lasting a number of $e$-folds of $\Delta N_1$ and $\Delta N_2$ respectively, the combined average equation of state is then simply
\beq
\bar{w} = \frac{\Delta N_1}{\Delta N_1+\Delta N_2}\bar{w}_1 + \frac{\Delta N_2}{\Delta N_1 + \Delta N_2} \bar{w}_2~.
\eeq
For our purposes, period 1 is between the end of inflation and the end of simulation time $t_e$, and period 2 is between the end of simulation time and the end of reheating. Between the end of inflation $t_{\rm end}$ and the end of simulation time $t_e$, the average equation of state $\bar{w}_1$ could be obtained by numerically evaluating the integral in \Eq{eq:wint_gen}. We can also obtain the number of $e$-folds $\Delta N_1$ from simulation results.

Between the end of simulation time $t_e$ and the end of reheating $t_{\rm re}$, an analytical estimate is needed based on what we have learned from the simulations. For tachyonic resonance without spillway {or spillway with a small Yukawa coupling $y^2/(8\pi)\lesssim 10^{-3}$}, numerical simulation shows that the equation of state quickly reaches a plateau and stays roughly a constant for the rest of the simulation. We will assume that the plateau behavior of the equation of state persists between the end of simulation time until the end of reheating. Therefore, the average equation of state during this period is simply $\bar{w}_2 = w_{\rm re}(t=t_e)$, the value at the end of simulation time. The number of $e$-folds $\Delta N_2$ between the end of simulation time and the end of reheating is also straightforward to obtain. From simulation, we know the value of the Hubble constant at the end of the simulation, $H_e = H(t=t_e)$. On the other hand, the Hubble constant at the end of reheating is defined by $H = \Gamma_{\phi}$. Therefore, the number of $e$-folds is simply
\beq
\Delta N_2 = \log\left(\frac{a(t=t_{\rm re})}{a(t=t_e)}\right) = \frac{2}{3(1+\bar{w}_2)}\log\left(\frac{\Gamma_{\phi}}{H_e}\right)~.
\eeq

The analytical estimate is slightly more involved for the case of {efficient spillway with a larger Yukawa coupling $y^2/(8\pi)> 10^{-3}$}, where the equation of state is observed to have a non-trivial time-dependence. In particular, for all simulations with efficient spillway turned on, the equation of state is observed to become redshift-dominated before the end of simulation time. In other words, the matter-like and radiation-like fluids at the end of the simulation are non-interacting, and we expect them to stay non-interacting until the end of reheating. We denote the equation of state observed at the end of simulation time as $w_e$. From $w_e$, we know
\beq
w_e = 0\times \frac{\rho_{{\rm m}, e}}{\rho_{{\rm m},e}+\rho_{{\rm r},e}}+\frac{1}{3}\times \frac{\rho_{{\rm r}, e}}{\rho_{{\rm m},e}+\rho_{{\rm r},e}}~,
\eeq
where $\rho_{{\rm m}, e}$ and $\rho_{{\rm r}, e}$ are the matter-like and radiation-like energy densities at the end of the simulation time, respectively.

Since the two fluids are non-interacting, they simply redshift as time evolves, and the equation of state at some later time is then
\beq
w(\Delta N) = 0\times \frac{\rho_{{\rm m}, e}}{e^{3\Delta N}}\frac{1}{\frac{\rho_{{\rm m}, e}}{e^{3\Delta N}}+\frac{\rho_{{\rm r}, e}}{e^{4\Delta N}}} + \frac{1}{3}\times \frac{\rho_{{\rm r}, e}}{e^{4\Delta N}}\frac{1}{\frac{\rho_{{\rm m}, e}}{e^{3\Delta N}}+\frac{\rho_{{\rm r}, e}}{e^{4\Delta N}}} = \frac{w_e}{e^{\Delta N}(1-3 w_e)+3 w_e}~,
\eeq
where $\Delta N$ is the number of $e$-folds between the end of simulation time to some later time of interest.

The average equation of state from $t_e$ until some later time is then
\beq
\bar{w} = \frac{1}{3}-\frac{\log\left(3 w_e + e^{\Delta N}(1-3 w_e)\right)}{3\Delta N}~.
\eeq
As a sanity check, when $w_e = 0$ or $w_e = 1/3$, $\bar{w}$ is identically $0$ or $1/3$ for all $\Delta N$, as it should be.

Now we need to find $\Delta N_2$ between the end of simulation time $t_e$ and the end of reheating $t_{\rm re}$. The ratio of total energy densities is the square of the ratio of the Hubble constants,
\beq
\frac{\rho_{\rm re}}{\rho_e} = \frac{\Gamma_{\phi}^2}{H_e^2}~.
\eeq
On the other hand, the ratio of energy densities can be computed from the way the energy density components redshift,
\beq
\begin{split}
\frac{\rho(\Delta N)}{\rho_e} &= \frac{\rho_{\rm m}(\Delta N)+\rho_{\rm r}(\Delta N)}{\rho_{{\rm m},e} + \rho_{{\rm r},e}} = \frac{\rho_{{\rm m},e}e^{-3 \Delta N}+\rho_{{\rm r},e}e^{-4\Delta N} }{\rho_{{\rm m},e} + \rho_{{\rm r},e}}\\
&= (1-3w_e)e^{-3 \Delta N}+3w_e e^{-4\Delta N}~. 
\end{split}
\eeq
From this equation, given $H_e$, $\Gamma_{\phi}$, and $w_e$, $\Delta N_2$ between the end of inflation and perturbative decay of the inflaton can be solved for, $\bar{w}_2$ can be determined, and the weighted sum of $\bar{w}_1$ and $\bar{w}_2$ can be calculated to obtain $\bar{w}_{\rm re}$.

\begin{figure}[H]
    \centering
\includegraphics[width=\textwidth]{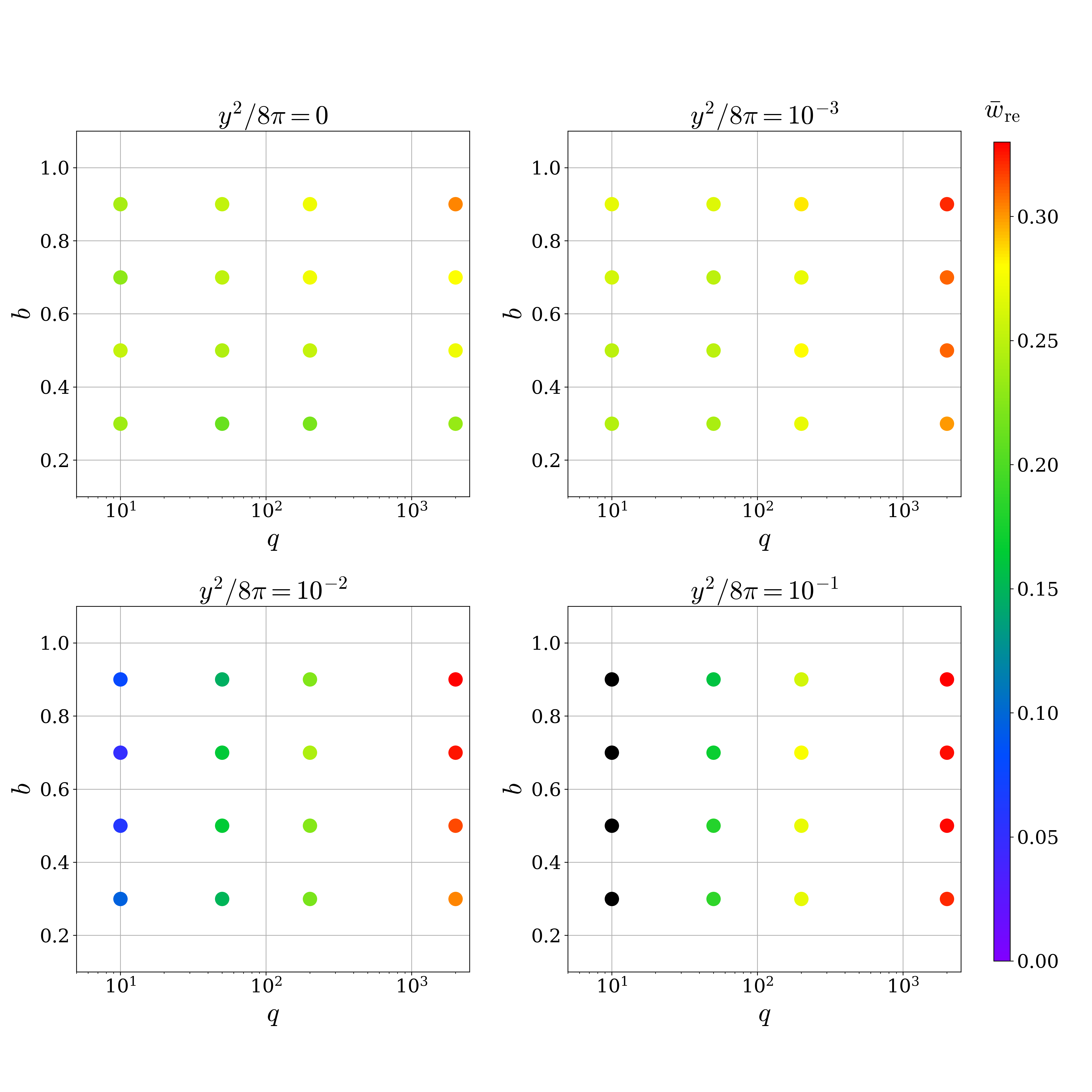}
    \caption{Estimates of $\bar w_{\rm re}$ against $q$ and $b$ for different choices of $y$. The black dots in the panel of $y^2/(8\pi) = 0.1$ indicate $\bar w_{\rm re}=0$.}
    \label{fig:wbar}
\end{figure}

{$\bar w_{\rm re}$ against $q$ and $b$ for four different choices of $y$ are presented in \Fig{fig:wbar}, using the computation procedure discussed above. Compared to the peak values of $w_{\rm re}$ shown in \Fig{fig:wspace2}, the most noticeable change is that in the intermediate range of $q$ ($10\lesssim q \lesssim 1000$), $\bar w_{\rm re}$ is reduced from the corresponding peak value, due to the redshift domination and the smaller equation of state between $t_e$ and $t_{\rm re}$. In addition, $\bar w_{\rm re}$ for the efficient spillway case with a larger $y$ is comparable or even smaller that of the tachyonic case in this range. It stems from different treatments of $w_{\rm re}$ between $t_e$ and $t_{\rm re}$ for the two cases. Strictly speaking, assuming that $w_{\rm re}$ remains constant after $t_e$ in the tachyonic case is a simplification and might overestimate $\bar{w}_{\rm re}$. It is beyond the scope of the paper to extend $t_e$ to achieve a more precise computation of the evolution of $w_{\rm re}$ in the tachyonic scenario. Another new feature appears for $y^2/(8\pi)=0.1$: $\bar{w}_{\rm re}$ demonstrates a clear increase as $q$ increases while the peak value stays close to $1/3$ for $q>10$. In this case, as $q$ increases, perturbative decays become more efficient, reducing the time between preheating and reheating. As a result, there is less time for $w_{\rm re}$ to decrease after the peak value is reached, causing $\bar{w}_{\rm re}$ to be larger.}

%%%%%%%%%%%%%%%%%%%%%%%%%%%%%%%%%%%%%%%%%
\subsection{Constraints on $n_s$ and $r$}
\label{subsec:model1res}
%%%%%%%%%%%%%%%%%%%%%%%%%%%%%%%%%%%%%%%%%

In the previous section, we have shown that turning on the spillway mechanism has two major effects on the average equation of state $\bar{w}_{\rm re}$ between the end of inflation and the end of reheating: at a low $q \sim 10$, $\bar{w}_{\rm re}$ is around 0.2 for tachyonic resonance, while more efficient spillway makes $\bar{w}_{\rm re}$ closer to 0. At a high $q$, on the other hand, turning on efficient spillway increases $\bar{w}_{\rm re}$ to the radiation-dominated value $1/3$. In this section, we discuss how different $\bar{w}_{\rm re}$'s are reflected in the inflationary observables $n_s$ and $r$. As representative examples, we consider three groups of inflation models: power-law potential, quartic hilltop, and $\alpha$-attractor. We will directly present the results first: the definitions of the inflation potentials and the relevant formula used could be found later in this section.

\begin{figure}[h]
    \centering
    \includegraphics[width=15cm]{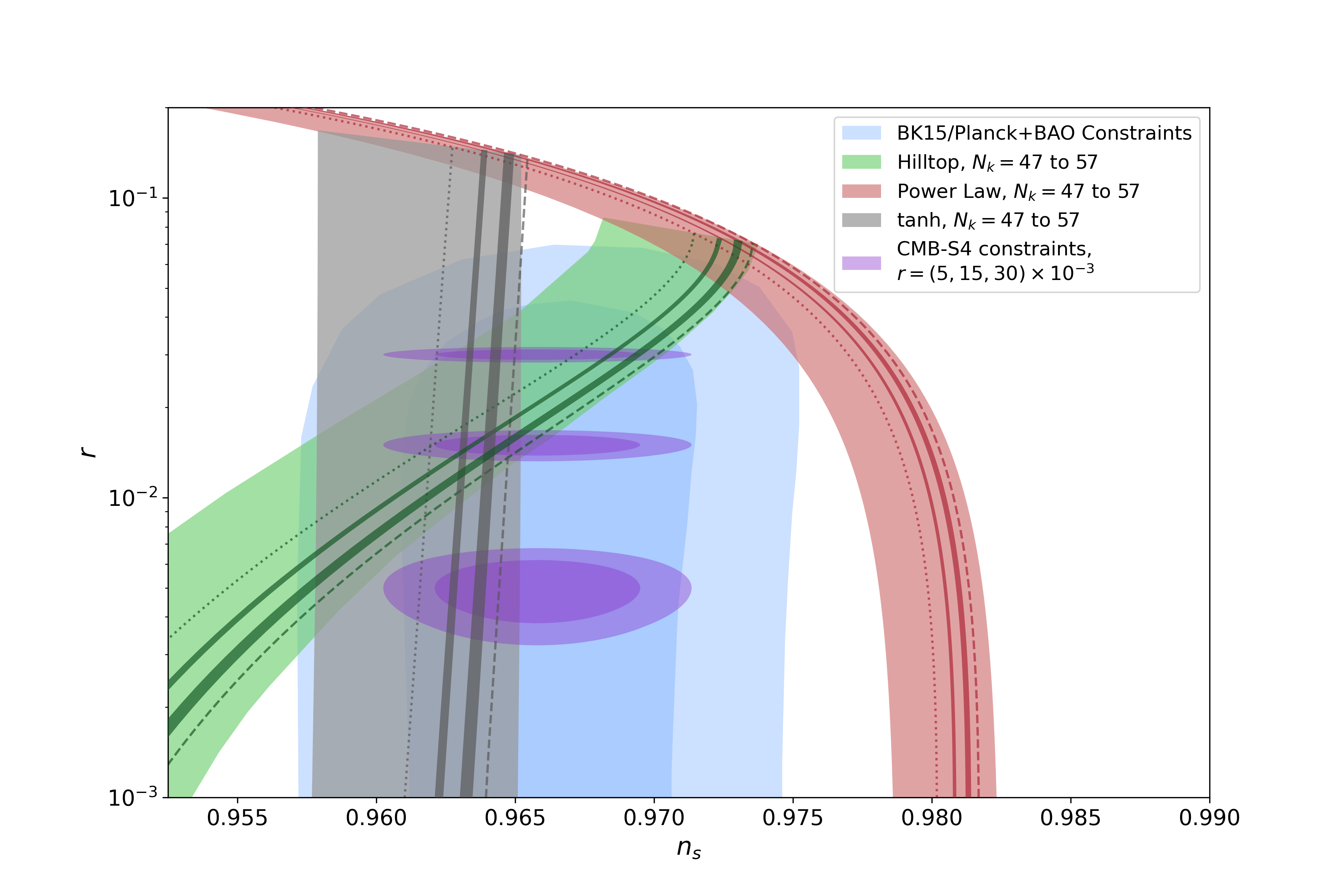}
    \caption{Constraints on $n_s$ and $r$ for various inflation models with different values of $\bar w_{\rm re}$ and $q$.  The blue region indicates the current $1\sigma$ and $2\sigma$ constraints on $n_s$ and $r$ \cite{nsr}, while the purple region shows projected CMB-S4 constraints \cite{nsr2} in the ($n_s$, $r$) plane. The light red, green, and gray regions denote ranges of predicted values in the same plane for the power-law, quartic hilltop, and tanh inflationary potentials respectively, fixing $N_k$ between 47 and 57. The dimensionless parameter $\alpha$ in all the models are varied between $0$ and $\infty$. The darker bands show the constraints imposed by spillway preheating for different regions of the parameter space: the dotted line is for inefficient spillway preheating with $\bar w_{\rm re}=0$ and $q=10$; the next one to the right is for $\bar w_{\rm re}=0.1$, $q\in [10,50]$; the next is $\bar w_{\rm re}=0.2$, $q\in [10,2000]$; the dashed line corresponds to $\bar w_{\rm re}=1/3$. 
    Estimates of the allowed range of $q$ for each $\bar w_{\rm re}$ are derived from \Fig{fig:wbar} assuming $y^2/(8\pi)=0.1$. 
    }
    \label{fig:nsr}
\end{figure}

The final results are presented in \Fig{fig:nsr}. It shows predicted values of $n_s$ and $r$ for the three representative inflation models overlaid on the current and near-future projected observational constraints. The light regions indicate constraints from fixing $N_k$ in a certain range (without specifying the preheating models), which commonly appear in the literature,  while the darker regions indicate the constraints imposed by fixing various values of $\bar{w}_{\rm re}$ that could be achieved in different regions of parameter space in the spillway preheating model. For a given $\bar{w}_{\rm re}$, the uncertainty of the prediction shrinks considerably. We will focus on the curves associated with $\bar{w}_{\rm re}>0$. 
We could see that when the spillway mechanism is most efficient with $\bar{w}_{\rm re}\approx 1/3$, there is almost a definite relation between $n_s$ and $r$ for a given class of inflation model with varying model parameters, represented by the dashed dark lines. 
In general, more efficient reheating (i.e. average equation of state closer to $\bar{w}_{\rm re} = 1/3$) leads to a larger $n_s$, corresponding to a bluer spectrum. For models that are on the edge of being excluded by Planck 2018 such as the power-law potential we considered here, turning on the efficient spillway with a high $q$ could push the model completely out of the $2\sigma$ constraint by Planck 2018. On the other hand, for inflation models that lean on the redder side for $n_s$ such as the hilltop and $\alpha$-attractor model, the spillway preheating is currently within the Planck 2018 constraint. As the sensitivities improve with the future CMB observations such as CMB-S4~\cite{nsr2}, an efficient preheating mechanism such as the spillway with a bluer spectrum would be more preferred if a smaller $r$ is observed, i.e., $r = (5, 15) \times 10^{-3}$ associated with the lowest two purple ellipse in the figure. 

For all inflation potentials, the general strategy for computing the predicted $n_s$ and $r$ values is the following: given an inflation potential, we can write the slow-roll parameters $\epsilon_V$ and $\eta_V$ at the time of horizon exit as a function of $\phi_k$, the inflaton field value at the time of horizon exit. Using the definition of $n_s$ and $r$ in terms of the slow-roll parameters \Eq{eq:nsslowroll}, \Eq{eq:rslowroll}, we obtain a relationship between $\phi_k$ and $n_s$/$r$. 
Similarly, one could express $\phi_e$, the inflaton field value at the end of inflation satisfying $\epsilon_V(\phi_e)=1$, in terms of $n_s$ and $r$. 
Plugging in the expression for $\phi_k$ and $\phi_e$ in terms of $n_s$ and $r$ into \Eq{eq:Nkslowroll}, we get an expression for $N_k$ in terms of the inflationary observables, $n_s$ and $r$ only, eliminating $\phi$'s. On the other hand, \Eq{eq:Nkconstraint} relates $N_k$ to reheating parameters $N_{\rm re}(\bar{w}_{\rm re})$ and $\bar{w}_{\rm re}$ (and also the inflationary observables through the ratio $\frac{V(\phi_k)}{V(\phi_e)}$). Equating the two expressions for $N_k$ results in a curve in $n_s$-$r$ space that varies as we change $\bar{w}_{\rm re}$. Below we collect the relevant equations for the three inflation models we have considered.

\hspace{0pt}

\noindent \underline{Model 1: Power law} Power-law inflation potential is given by
\begin{align}
    V(\phi)=\frac12 m^{4-\alpha}\phi^{\alpha}~,
\end{align}
for some real number $\alpha$.  This general potential encompasses several specific inflation models of interest: Setting $\alpha=2$  reproduces ``$m^2\phi^2$" inflation \cite{LINDE1983177, Belinsky:1985zd, PIRAN1985331}, and axion monodromy inflation \cite{McAllister_2010, Silverstein_2008} suggests $\alpha=2/3$ or $\alpha=1$. The corresponding slow-roll parameters are given by
\begin{align}
    \label{eq:powerepsilon}
    \epsilon_V&=\frac{M_{\text{pl}}^2}{2}\left(\frac{V'}{V}\right)^2=\frac{M_{\text{pl}}^2}{2}\alpha^2\phi^{-2}_k~,\\
    \label{eq:powereta}
    \eta_V&= M_{\text{pl}}^2\frac{V''}{V}=M_{\text{pl}}^2\alpha(\alpha-1)\phi^{-2}_k~,
\end{align}
Eliminating $\phi_k$ from $n_s$, $r$, and $N_k$,\footnote{Here we ignore the negligible contribution from $\phi_e$ in Eq. (\ref{eq:Nkslowroll}).} we obtain
\begin{align}
    1-n_s&=\frac{\alpha+2}{2N_k}~, \quad r=8\alpha\frac{1-n_s}{\alpha+2}~.\label{eq:powerr}
\end{align}
On the other hand, $N_k$ is given by Eq.~\eqref{eq:Nkconstraint}, which depends on the inflation potential through the ratio~\cite{Dai:2014jja}
\begin{align}
    \label{eq:powerVk}
    \frac{V(\phi_k)}{V(\phi_{\text{e}})}=\left(\frac{16}{r}\right)^{\alpha/2}~,
\end{align}
and the reheating parameters $N_{\rm re}$ and $\bar{w}_{\rm re}$. Numerically evaluating \Eq{eq:Nkconstraint}, \Eq{eq:powerr}, and \Eq{eq:powerVk} yields the result plotted in red in Figure~\ref{fig:nsr}.

\hspace{0pt}

\noindent \underline{Model 2: Quartic hilltop} The quartic hilltop potential is a special case of the hilltop potential introduced by \cite{Boubekeur_2005} that is consistent with latest observations from Planck 2018 \cite{Akrami:2018odb}. The potential takes the form
\begin{align}
    \label{eq:quarticV}
    V=V_0\left(1-\alpha\left(\frac{\phi}{M_{\text{pl}}}\right)^4\right)~,
\end{align}
where $\alpha$ is a dimensionless real constant. We compute the slow-roll parameters to be
\begin{align}
    \label{eq:hilltopepsilon}
    \epsilon_V=\frac{M_{\text{pl}}^2}{2}\left(\frac{-4\alpha\frac{\phi_k^3}{M_{\text{pl}}^4}}{1-\alpha\frac{\phi_k^4}{M_{\text{pl}}^4}}\right)^2~, \quad
    \eta_V=M_{\text{pl}}^2\frac{-12\alpha\frac{\phi_k^2}{M_{\text{pl}}^4}}{1-\alpha\frac{\phi_k^4}{M_{\text{pl}}^4}}~.
\end{align} 
The explicit formula for $\phi_k$ has been worked out in \cite{Germ_n_2021}. Eliminating $\phi_k$ from the definition of $n_s$, $r$, and $N_k$ in \Eq{eq:nsslowroll}, \Eq{eq:rslowroll}, and \Eq{eq:Nkslowroll}, we get
\begin{align}
    N_k=\frac{24\left(8 (1-n_s)-\sqrt{3 r\left(16 (1-n_s)-3 r\right)}\right)}{\left(8 (1-n_s)-3 r\right)^{2}}~.
\end{align}
Equating this expression to the alternative definition of $N_k$ in \Eq{eq:Nkconstraint}, we obtain the results plotted in green in \Fig{fig:nsr}.

\hspace{0pt}

\noindent \underline{Model 3: Tanh potential} Lastly, we consider a specific form of $\alpha$-attractor potential \cite{Kallosh:2013hoa}
\begin{align}
    \label{eq:tanh}
    V(\phi)=V_0\tanh^2\left (\frac{\phi}{\alpha M_{\text{pl}}}\right )~,
\end{align}
where $V_0$ is a dimensionful constant while the other constant $\alpha$ is dimensionless. The slow-roll parameters are
\begin{align}
    \epsilon_V&=\frac{8}{\alpha^2}\csch^2\left(\frac{2\phi_k}{\alpha M_{\text{pl}}}\right)\label{eqn:evtanh}~,\\
    \eta_V&=\frac{2}{\alpha^2}\left(4\csch^2\frac{2\phi_k}{\alpha M_{\text{pl}}}-2\sech^2\frac{\phi_k}{\alpha M_{\text{pl}}}\right)~.
\end{align}
Eliminating $\phi_k$, we obtain
\begin{align}
    N_k&=\frac{\alpha^2}{8}\left(\sqrt{\frac{128}{\alpha^2 r}+1}-\sqrt{\frac{8}{\alpha^2}+1}\right)~.
\end{align}
On the other hand, \Eq{eq:Nkconstraint} expresses $N_k$ in terms of the reheating parameters $N_{\rm re}$ and $\bar{w}_{\rm re}$ and the ratio
\beq
\frac{V(\phi_k)}{V(\phi_e)}=\frac{16(\sqrt{\alpha^2+8}+\alpha)^2}{\alpha^2 r+128}~.
\eeq
Equating the two expression for $N_k$, we obtain the result plotted in gray in \Fig{fig:nsr}.

%%%%%%%%%%%%%%%%%%%%%%%%%%%%%%%%%%%%%%%%%
\section{Observable 2: Gravitational Waves}
\label{sec:obs2}
%%%%%%%%%%%%%%%%%%%%%%%%%%%%%%%%%%%%%%%%%
The second observable effect of preheating we study is the stochastic gravitational wave background (SGWB) spectrum. In this section, we simulate the SGWB in the spillway scenario and examine whether and how the spillway mechanism could alter the SGWB for tachyonic preheating as studied in~\cite{Figueroa:2017vfa}.

%%%%%%%%%%%%%%%%%%%%%%%%%%%%%%%%%%%%%%%%%
\subsection{High-frequency Gravitational Waves}
\label{subsec:gwdef}
%%%%%%%%%%%%%%%%%%%%%%%%%%%%%%%%%%%%%%%%%

In general, fragmentation of the inflaton condensate in the early stages of preheating generates a quadrupole moment in the matter distribution which sources high-frequency GWs (see \cite{Amin:2014eta, Lozanov:2019jxc} for reviews). In this section, we provide a review of the dimensional analysis estimate of the frequency and amplitude of the resulting GWs from topological defects, relations between the GWs today and those during preheating, and the contribution from GWs to the effective number of relativistic degrees of freedom. These are general discussions, applying to broad classes of preheating scenarios. Readers who are interested in the specific results for spillway preheating from simulations could jump to Sec.~\ref{subsec:model2res}.

\begin{figure}[h]
    \centering
    \includegraphics[width=16cm]{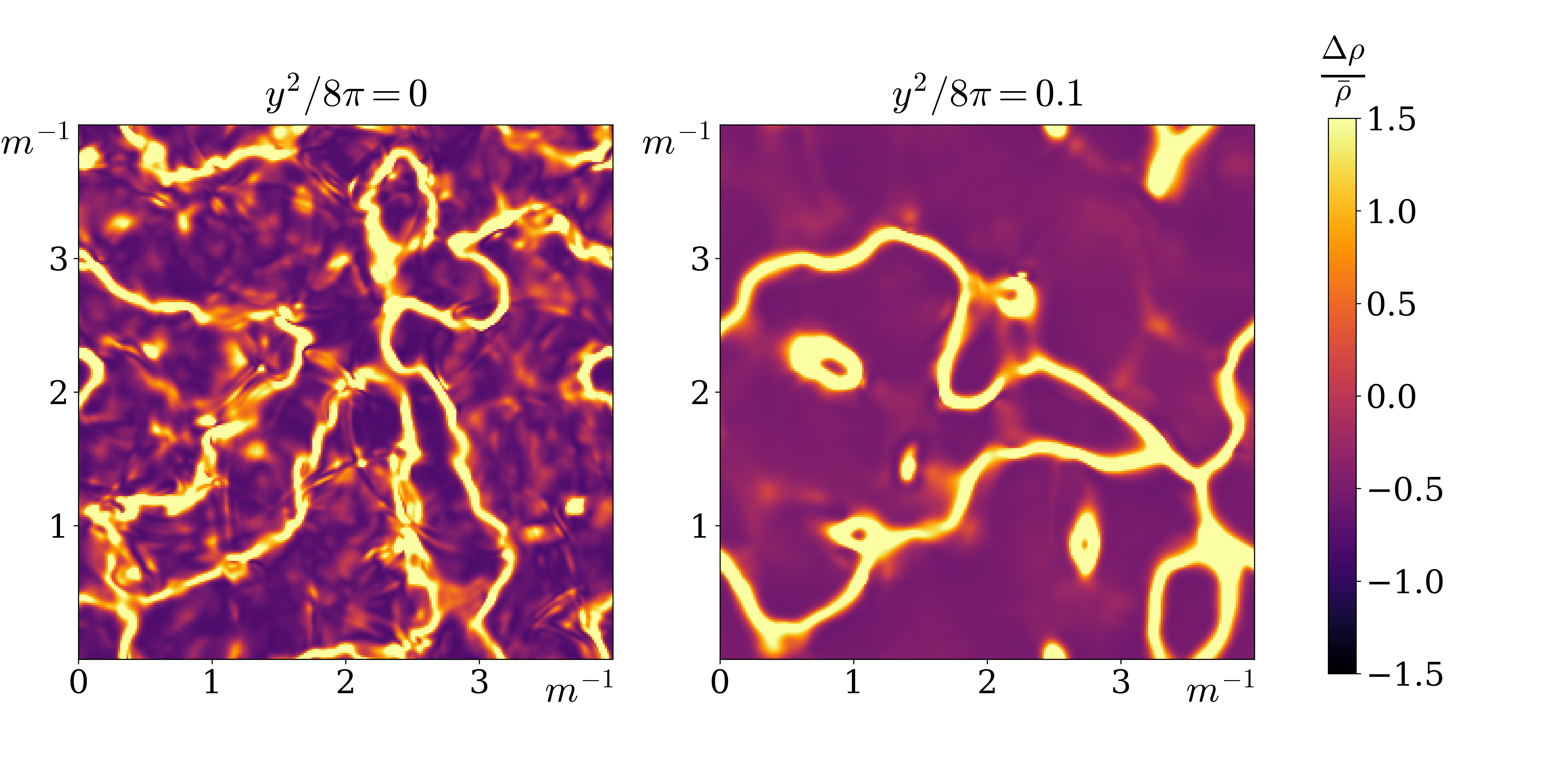}
    \caption{Snapshot of a two-dimensional slice of the deviation of total energy density $\Delta\rho$ from the average $\bar{\rho}$ at $mt=11$. Images correspond to tachyonic resonance (left) and spillway preheating (right) for $q=200$, $b=0.9$.}
    \label{fig:bubble}
\end{figure}

Our discussion mainly follows that of ~\cite{PhysRevD.77.043517}. The main source of GW production comes from topological defects formed, i.e., domain walls in the scalar system we consider. This could be estimated by considering a spherical bubble with radius $R$ and quadrupole moment $Q$, which emits GW with a power~\cite{PhysRevD.77.043517}
\begin{align}
    P_{\rm gw, g}\sim G \dddot Q^2~, 
\end{align}
where the subscript g denotes quantities at the time of generation. The quadruple moment that generate the gravitational wave is sourced by the $\chi$ energy density at the time of generation,
\beq
Q_{ij} = \int \dd[3]{x} \left(x_i x_j - \frac{x^2 \delta_{ij}}{3}\right)\rho_{\chi, {\rm g}}~,
\eeq
which we can approximate as $Q\sim \rho_{\chi, {\rm g}}R^5$. Then the total power of GW could be estimated to be \begin{align}
    P_{\rm gw,  g}\sim G \left[\dv[3]{t} (\rho_{\chi,{\rm g}} R^5) \right]^2\sim G\rho_{\chi,{\rm g}}^2 R^4~.
\end{align}
Since $P_{\rm gw, g}\sim\rho_{\rm gw, g}R^2$, $\rho_{\phi, {\rm g}}\sim \phi_{\rm g}^2 m^2$, and from simulations with one benchmark shown in \Fig{fig:bubble}, $R\sim 10 m^{-1}$,  we find the GW energy density to be
\begin{align}
    \frac{\rho_{\rm gw, g}}{\rho_{\chi, {\rm g}}}\sim G\rho_{\chi,{\rm g}} R^2\sim\frac{\rho_{\chi, {\rm g}}}{M_{\text{pl}}^2}\frac{10}{m^2}=\frac{\rho_{\chi, {\rm g}}}{\rho_{\phi, {\rm g}}}\frac{10\rho_{\phi,{\rm g}}}{M_{\text{pl}}^2m^2}\sim 10\frac{\rho_{\chi,{\rm g}}}{\rho_{\phi,{\rm g}}}\frac{\phi_{\rm g}^2}{M_{\text{pl}}^2}~,
\end{align}
where we approximate $G \sim 1/(10 M_{\rm pl}^2)$.
The fractional energy density in GWs is then
\begin{align}
    \Omega_{\rm gw, g}\sim\frac{\rho_{\rm gw, g}}{\rho_{\rm g}}\sim \frac{\rho_{\chi,{\rm g}}}{\rho_{\rm g}}\frac{10\rho_{\chi,{\rm g}}}{\rho_{\phi,{\rm g}}}\frac{\phi_{\rm g}^2}{M_{\text{pl}}^2}\sim 10^{-4}\text{\textendash} 10^{-2}\label{eq:topo}~,
\end{align}
where $\rho_{\rm g}$ is the total energy density stored in all the scalar fields at the time of GW production, and we approximate various energy density ratios with values that are consistent with simulation results: $\rho_{\chi, {\rm g}} /\rho_{\phi, {\rm g}}\sim \left(0.01\text{\textendash} 1\right)\rho_{\phi,{\rm g}}/\rho_{\rm g}$, $\left(\frac{\rho_{\chi, {\rm g}}}{\rho_{ {\rm g}}}\frac{\rho_{\phi,{\rm g}}}{\rho_{\rm g}}\right)\sim 0.1$, $\phi_g \sim 0.1 M_{\text{pl}}$.

After being produced, the GW redshifts following the evolution of the universe. 
In practice, we could use the GW spectrum by the end of the simulation with the scale factor $a_e$ and the Hubble parameter $H_e$, and obtain the present-day peak frequency by incorporating the redshifts between the end of the simulation and today: 
\begin{align}
    f_{0}= f_e \; \frac{a_e}{a_0} = \frac{k_e}{2\pi H_e}\frac{a_{e}}{a_{\rm RD}}\frac{a_{\rm RD}}{a_0}H_e~,
\end{align}
where $f_e = k_e/(2 \pi)$ with $k_e$ the corresponding wave number, and $a_{\rm RD}$ in between $a_e$ and $a_0$ is the scale factor when the universe becomes fully radiation dominated. 
We can rewrite the expansion after radiation domination using the conservation of comoving entropy density and temperature-radiation energy density relation:
\begin{align}
    \frac{a_{\rm RD}}{a_{0}}=\left(\frac{g_{s,0}}{g_{s, {\rm RD}}}\right)^{1/3}\frac{T_{0}}{T_{\rm RD}}=\left(\frac{g_{s,0}}{g_{s, {\rm RD}}}\right)^{1/3}\left(\frac{g_{\rm RD}}{g_{0}}\right)^{1/4}\left(\frac{\rho_{\rm rad,0}}{\rho_{\rm RD}}\right)^{1/4}= G_{\rm RD}\left(\frac{\rho_{\rm rad,0}}{\rho_{\rm RD}}\right)^{1/4}~,
\end{align}
where $\rho_{{\rm rad},0}$ is the radiation energy density today, $\rho_{\rm RD}$ is the total energy density when the universe becomes radiation dominated, and $G_{\rm RD} \equiv \left({g_{s,0}}/{g_{s, {\rm RD}}}\right)^{1/3}\left({g_{\rm RD}}/{g_{0}}\right)^{1/4}$ with $g_s$ the number of degrees of freedom associated with entropy and $g$ the number of degrees of freedom associated with energy density. 
Next we break $H_e$ into two factors of $H_e^{1/2}$, and rewrite one of them using the Friedmann equation
\begin{align}
H_e&=\sqrt{H_e}\sqrt{H_e}=\sqrt{H_e}\left(\frac{\rho_e}{3}\right)^{1/4}\frac{1}{\sqrt{M_{\text{pl}}}}=\sqrt{\frac{H_e}{M_{\text{pl}}}}\left(\frac{1}{3} \right)^{1/4}\rho_e^{1/4}~,
\end{align}
where the critical energy density by the end of the simulation is given by $\rho_e = 3 H_e^2 M_{\rm pl}^2$. 
Between $a_e$ and $a_{\rm RD}$, the equation of state is $\bar{w}_2$ and the total energy density evolves as $a^{-3(1+\bar{w}_2)}$. We can then rewrite
\begin{align}
    \frac{a_e}{a_{\rm RD}}\left(\frac{\rho_e}{\rho_{\rm RD}}\right)^{1/4}=\left(\frac{a_e}{a_{\rm RD}}\right)^{1+\frac{-3-3\bar{w}_2}{4}}=\left(\frac{a_e}{a_{\rm RD}}\right)^{\frac{1-3\bar{w}_2}{4}}= \epsilon_e^{1/4}~,
\end{align}
where $\epsilon_e \equiv (a_e/a_{\rm RD})^{1-3\bar{w}_2}$. Putting the equations above together, the peak frequency today is given by
\begin{align}
    f_0=\left(\frac13\right)^{1/4}\epsilon_e^{1/4} G_{\rm RD} \left(\frac{k_e}{H_e}\right)\left(\frac{H_e}{M_{\text{pl}}}\right)^{1/2}\frac{\rho_{\rm rad,0}^{1/4}}{2\pi}~.\label{eq:redshift2}
\end{align}

Now we proceed to calculate the shift in the amplitude as GWs travel through the universe. The GW energy density spectrum by the end of the simulation is defined as
\begin{align}
    \Omega_{{\rm GW},e}(f)=\frac{1}{\rho_{e}}\left(\dv{\rho_{{\rm GW}}}{\log k}\right)_e~.
\end{align}
Since GWs redshift as radiation, the present-day GW energy density satisfies $\rho_{\rm GW,0}/\rho_{{\rm GW}, e}\propto (a_0/a_e)^{-4}$, where $0$ subscripts denote today. Therefore the present-day spectrum is
\begin{align}
    \Omega_{\rm GW,0}(f)=\frac{1}{\rho_{c,0}}\left(\dv{\rho_{\rm GW}}{\log k}\right)_0=\left(\frac{a_e}{a_0}\right)^4\frac{1}{\rho_{c,0}}\left(\dv{\rho_{{\rm GW}}}{\log k}\right)_e=\left(\frac{a_e}{a_0}\right)^4\frac{\rho_{e}}{\rho_{c,0}}\Omega_{{\rm GW},e}(k)~,
\end{align}
where $\rho_{c,0}$ is the critical density today. 
In terms of the parameters $\epsilon_e$ and $G_{\rm RD}$ defined previously, the amplitude becomes 
\begin{align}
    \Omega_{\rm GW,0}(f)&=\epsilon_e G_{\rm RD}^4\frac{\rho_{{\rm rad},0}}{\rho_{c,0}} \Omega_{{\rm GW},e}(k)~, \nonumber \\
    &=\epsilon_e G_{\rm RD}^4\Omega_{{\rm rad},0}\Omega_{{\rm GW},e}(k)~,\label{eq:redshift}
\end{align}
where $\Omega_{{\rm rad},0}$ is the fraction of radiation energy density today.

Lastly, we can relate the peak amplitude of the SGWB spectrum to the number of effective degrees of freedom beyond the Standard Model. In the Standard Model, radiation energy density, $\rho_{\rm rad}$, consists of the contributions from photons and neutrinos. $N_{\text{eff}}$ then serves as a measure of the effective number of neutrino species: 
\begin{align}
\rho_{\rm rad}=\rho_\gamma+\rho_\nu=\rho_\gamma\left(1+\frac78\left(\frac4{11}\right)^{4/3}N_{\text{eff}}\right)~,
\end{align}
where $\rho_\gamma$ and $\rho_\nu$ are the energy densities of photon and neutrinos, respectively. The prefactor $(4/11)^{4/3}$ comes from the temperature ratio $T_\nu / T_\gamma =(4/11)^{1/3}$ between the photon and neutrino baths of the universe, and $(7/8)$ accounts for the Fermi distribution of neutrinos. GWs also contribute to the radiation energy density, and manifest as a correction term $\Delta N_{\text{eff}}$ to the effective degrees of freedom which we define as
\begin{align}
    \rho_{\rm GW}\equiv\rho_\gamma \times \frac78\left(\frac4{11}\right)^{4/3}\Delta N_{\text{eff}}~. 
\end{align}
Equivalently, we have
\begin{align} \label{eq:NeffGW}
    \Delta N_{\mathrm{eff}}=\frac{h_0^2 \Omega_{\mathrm{GW},0}}{h_0^2 \Omega_{\gamma, 0}} \; \frac{8}{7}\left(\frac{11}{4}\right)^{4 / 3}~,
\end{align}
where $\Omega_{\rm{GW},0}$ is an integration of $\Omega_{\rm{GW},0}(k)$ over the entire frequency range and the present-day photon energy density is known to be $h_0^2 \Omega_{\gamma,0}=2.47 \times 10^{-5}$ \cite{Workman:2022ynf}. The current Planck bound sets $\Delta N_{\mathrm{eff}}\leq 0.29$ \cite{Akrami:2018odb}, and the future CMB-S4 could improve the sensitivity to $\Delta N_{\mathrm{eff}}\leq 0.06$ \cite{nsr}. This could potentially act as another constraint on preheating models.

%%%%%%%%%%%%%%%%%%%%%%%%%%%%%%%%%%%%%%%%%
\subsection{Simulation Methods}
\label{subsec:model2sim}
%%%%%%%%%%%%%%%%%%%%%%%%%%%%%%%%%%%%%%%%%

To have a more precise determination of the GW spectrum, we need to rely on numerical simulations. We use the Cosmolattice library~\cite{Figueroa_2023} with a lattice size $512^3$. The IR and UV cutoffs are adjusted as necessary to capture the entire spectrum, varying between simulations. Due to the increased computational load from the larger lattice size, we only run simulations up to $t=50m^{-1}$. We find that majority of  GW production happens prior to this time, and further evolution could be characterized purely by the redshift analysis described in the previous section. Thus this earlier time cutoff should not affect the final results. In addition, the increased efficiency of Cosmolattice package allows us to maintain the energy conservation with a precision of $10^{-3}$ once the time step is refined to $0.004 m^{-1}$.

Below we will briefly outline the procedure of computations implemented by Cosmolattice package. The anisotropic stress tensor, $\Pi_{ij}$, could be written in terms of the stress-energy tensor $T_{ij}$ of the fields, the pressure $p$, and the metric $g_{ij}=a^2(t)(\delta_{ij}+h_{ij})$, as 
\begin{align}
    \label{eq:anisotropictensor}
    \Pi_{ij}\equiv T_{ij}-pg_{ij}~.
\end{align}
Cosmolattice computes the transverse-traceless part of this tensor, $\Pi^{TT}_{ij}=\Lambda_{ijlm}\Pi_{lm}$, with the projection operator defined in momentum space as
\begin{align}
    \label{eq:projector}
\Lambda_{i j l m}(\hat{\mathbf{k}}) &\equiv P_{i l}(\hat{\mathbf{k}})
P_{j m}(\hat{\mathbf{k}})-\frac{1}{2} P_{i j}(\hat{\mathbf{k}}) P_{l m}(\hat{\mathbf{k}})~, \\
\label{eq:momentumprojector}
P_{i j}&=\delta_{i j}-\hat{k}_i \hat{k}_j~,
\end{align}
where $\hat k_i$ denotes the unit momentum vector. The metric tensor $h_{ij}$ is then evolved according to the linearized Einstein equations on the FLRW background:
\begin{align}
    \label{eq:heom}
    \ddot{h}_{i j}+3 H \dot{h}_{i j}-\frac{\nabla^2}{a^2} h_{i j}=\frac{2}{M_{\text{pl}}^2 a^2} \Pi_{i j}^{\mathrm{TT}}~.
\end{align} 
However, computing the projection $\Pi_{ij}^{TT}$ at each time step is computationally expensive. Cosmolattice overcomes this limitation by decomposing $h_{ij}$ in momentum space as
\begin{align}
    \label{eq:u}
    h_{ij}(k)\equiv \Lambda_{ijlm}(k)u_{lm}(k)~,
\end{align}
and evolving $u_{lm}$ according to
\begin{align}
    \ddot{u}_{l m}+3 H \dot{u}_{l m}-\frac{\nabla^2}{a^2} u_{lm}=\frac{2}{M_{\text{pl}}^2 a^2} \Pi_{lm}~.
\end{align} 
We could then extract the value of $h_{ij}$ from the computed $u_{lm}$ using Eq.~\eqref{eq:u}.

%%%%%%%%%%%%%%%%%%%%%%%%%%%%%%%%%%%%%%%%%
\subsection{Results}
\label{subsec:model2res}
%%%%%%%%%%%%%%%%%%%%%%%%%%%%%%%%%%%%%%%%%

In this section, we will present the results of GW production in spillway preheating based on the numerical method described in the previous section, and compare them with those from tachyonic resonance preheating without the spillway mechanism.

\begin{figure}[h]
    \centering
    \includegraphics[width=7.8cm]{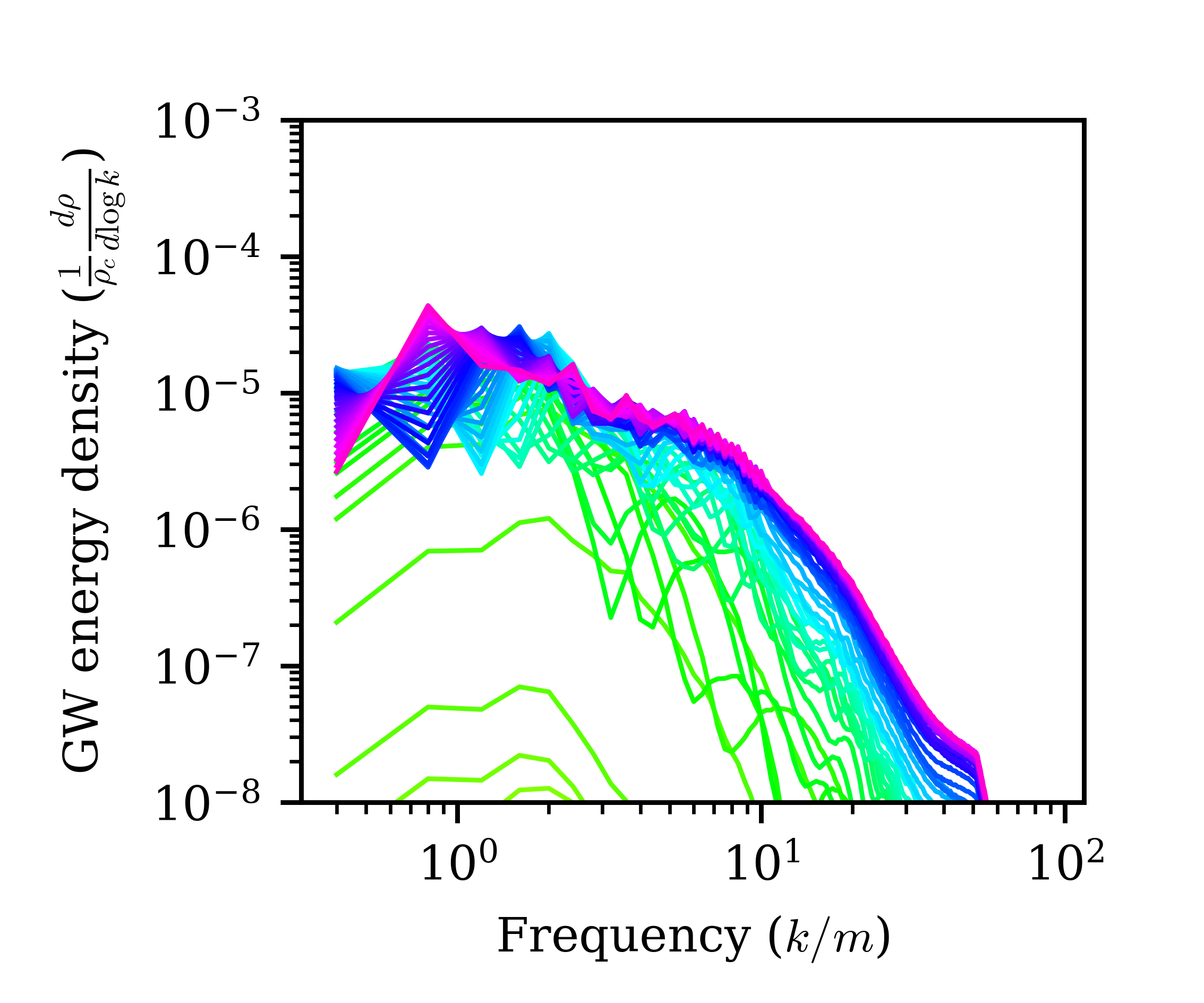}
    \includegraphics[width=7.0cm]{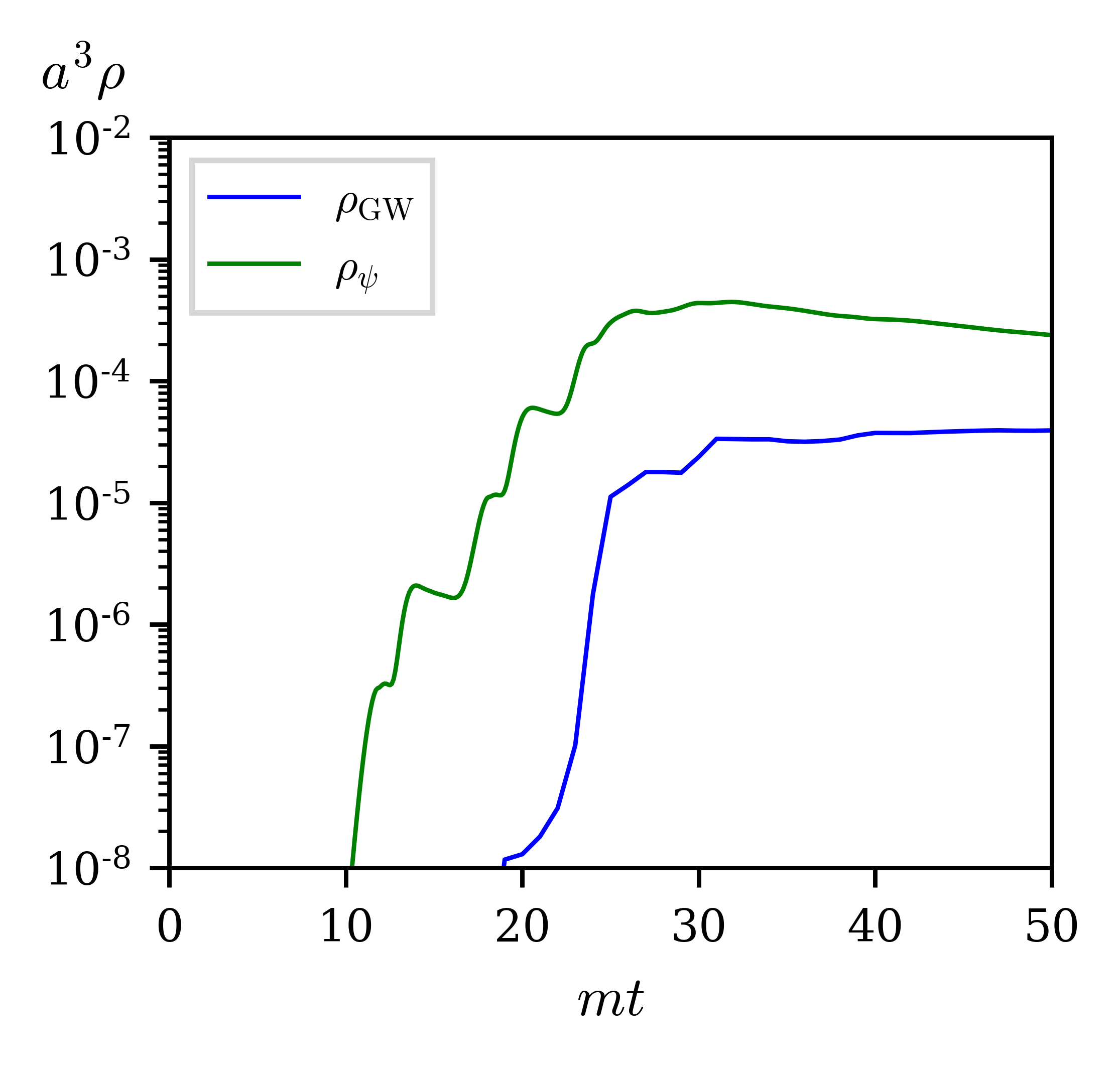}
    \caption{Left: SGWB spectrum generated from simulations for a spillway model with $q=30$, $b=0.5$, $y^2/8\pi=0.1$. Time evolves from green to pink. Right: time evolution of $\rho_{\rm GW}$ (blue) and $\rho_\psi$ (green).}\label{fig:gwspectrum}
\end{figure}

In \Fig{fig:gwspectrum}, we present a sample simulation of the spillway model to demonstrate some general features. This figure is based on $q=30$, $b=0.5$, $y^2/8\pi=0.1$ with GWs evolving in time from green to pink. As shown in the left panel of \Fig{fig:gwspectrum}, after $t \sim 40 m^{-1}$, the overall shape of the gravitational wave spectrum remains unchanged. The reason for the gravitational wave spectrum to stop evolving is displayed in the right panel of \Fig{fig:gwspectrum}, which shows that the energy density of GWs is roughly constant beyond $t=40 m^{-1}$. After this point, the evolution of GWs is dominated by the effect of redshift. This has been observed in all our simulations for $q>10$. At $q\leq 10$, it can take up until $mt=150$ for the GW energy density to plateau. In these simulations, the timestep in our simulation was further refined to $0.001m^{-1}$ to maintain energy conservation. Consequently, to reduce the computational costs, each simulation is only run until this plateau behavior is observed, and the GW spectrum today is computed analytically via redshifting the spectrum from the simulations using Eq.~\eqref{eq:redshift}. The right panel also shows that the decays of the scalar fields to fermions become effective roughly at the same time as GW production, though slightly earlier. In other words, the spillway is turned on about the same time as GWs are produced and could affect the production and evolution of the GWs, as we will explain more below.

\begin{figure}[h]
    \centering
    \includegraphics[width=7.5cm]{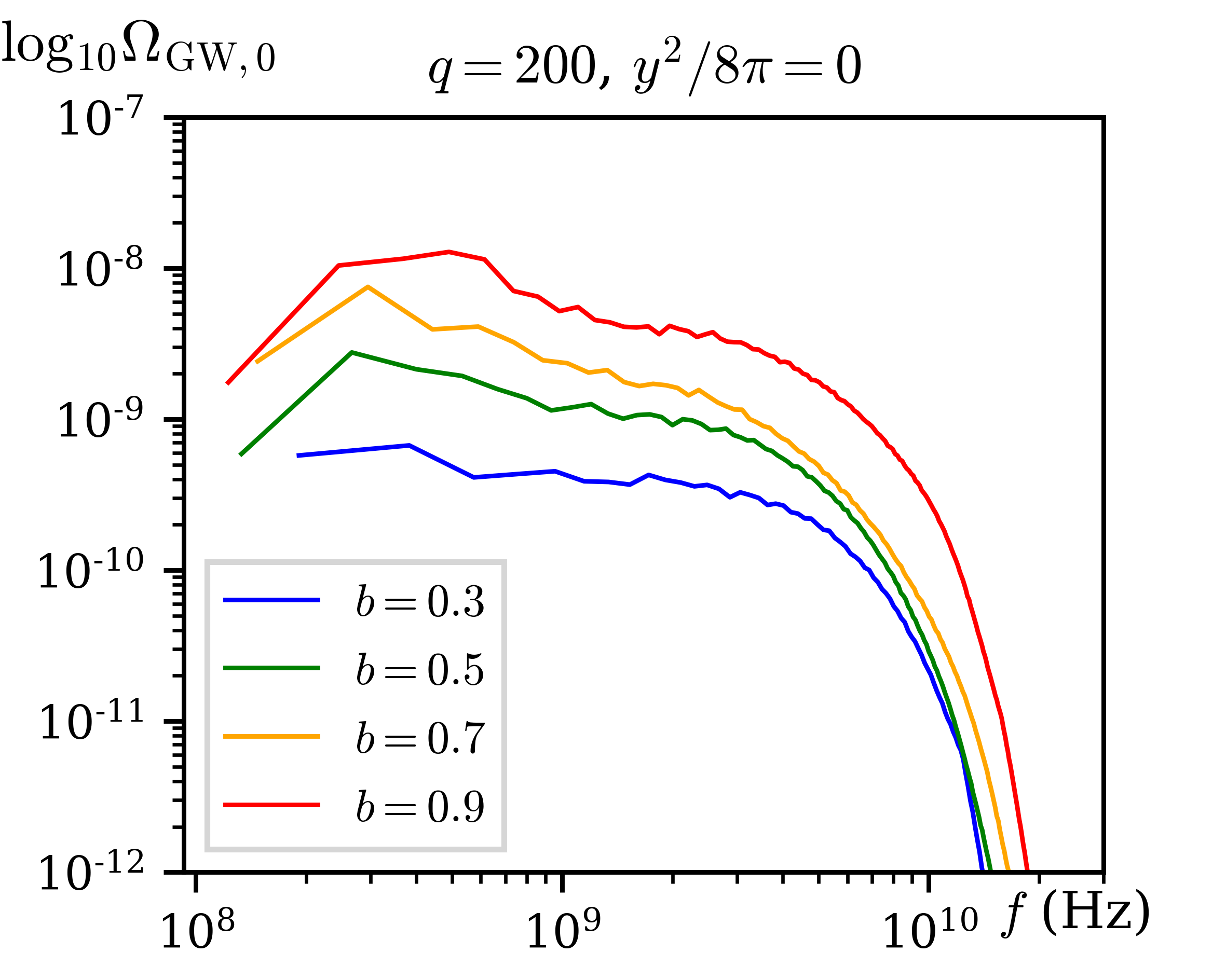}
    \includegraphics[width=7.5cm]{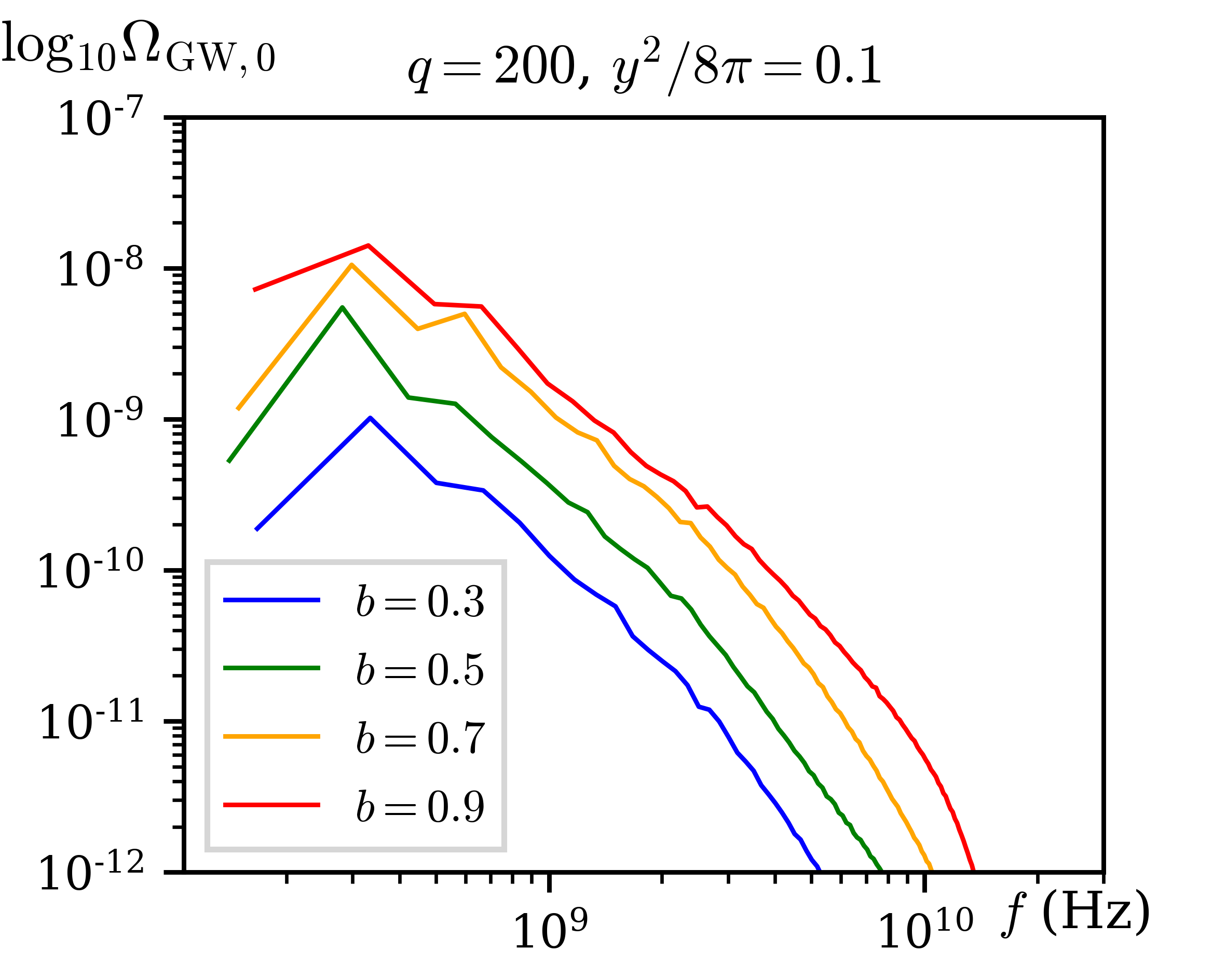}
    \caption{GW spectra today for models with $q=200$ and different $b$'s. Left: $y=0$; Right: $y^2/(8\pi) =0.1$. }
    \label{fig:q50} 
\end{figure}

The present-day GW spectra for various values of $b$ and two values of $y$ at a fixed $q=200$ are presented in \Fig{fig:q50}. From the figure, one could see that the GW amplitude increases as $b$ increases, for both tachyonic resonance and spillway preheating. As $b$ increases, more $\chi$ particles are produced via the tachyonic instabilities due to $\phi \chi^2$, leading to more field fragmentation and boosted GW production. On the other hand, the amplitudes of GWs at higher frequencies are considerably smaller with $y^2/8 \pi = 0.1$ compared to the case with $y=0$, for a given pair of $q$ and $b$. When spillway is efficient, the perturbative decays $\chi \to \bar\psi {\psi}$ erase topological defects in the $\chi$ energy density, as shown in Fig.~\ref{fig:bubble}, damping the higher-momenta modes more than the lower modes. Thus spillway models predict sharper peaks in the GW spectra.  These spillways could also change $\bar{w}_{\rm re}$ as discussed in Sec.~\ref{sec:obs1} and thus the following evolution of GWs. Yet numerically this turns out to be a minor effect.

\hspace{0pt}

\begin{figure}[h]
    \centering
    \includegraphics[width=16cm]{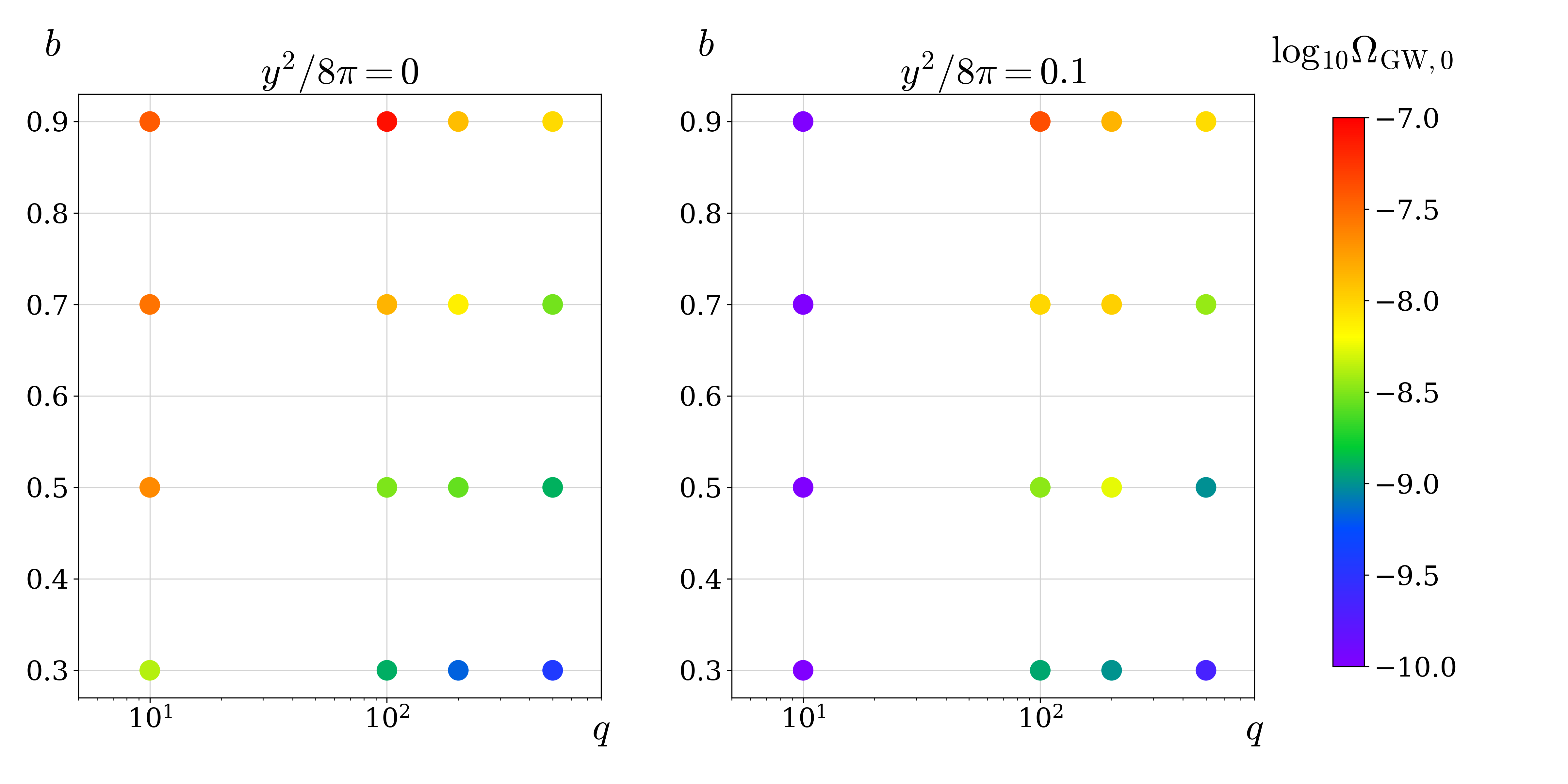}
    \caption{$\Omega_{{\rm GW},0}$ as a function of $q$ and $b$. Left: $y=0$; right: $y^2/(8\pi)=0.1$. }
    \label{fig:gw_peaks}
\end{figure}

The peak GW amplitude throughout the two-dimensional parameter space ($q - b$) is presented in \Fig{fig:gw_peaks}. Two trends are evident in the figure:~{\it 1)} for a given $q>10$, as $b$ increases, the peak amplitude increases, {\it 2)} for  $q \gtrsim 100$, as $q$ increases, the peak amplitude decreases. These reflect that $\Omega_{{\rm GW},0}$ should trace the maximum value of $\expval{\chi^2}$, which is proportional to $b$ and $\expval\phi$ and inversely proportional to $q$ as shown in Eq. \eqref{eq:chicrit}. We note that in general we expect $\expval{\phi}$ should grow when we increase $q$, so there may be cases where increasing $q$ still results in an increase in $\Omega_{\rm{GW},0}$ if the growth in $\expval{\phi}$ compensates the $q$ suppression. In \Fig{fig:gw_peaks}, this occurs when $10\lesssim q\lesssim 100$ and $b\sim 0.9$. This is consistent with the result in \Fig{fig:wspace2}, which shows that the equation of state is dominated by $\expval{\phi}$ in this regime.

These trends apply to both tachyonic resonance and spillway preheating. The peak amplitudes, however, do not depend strongly on the choices of $y$'s, except for the low $q$ region with $q \sim 10$. In this region, the presence of spillways actually suppresses the non-perturbative particle production, as explained in Sec.~\ref{subsec:model1sim}, leading to an even more suppressed GW production, compared to the $y=0$ case.

\begin{figure}[h] 
    \centering
    \includegraphics[width=11cm]{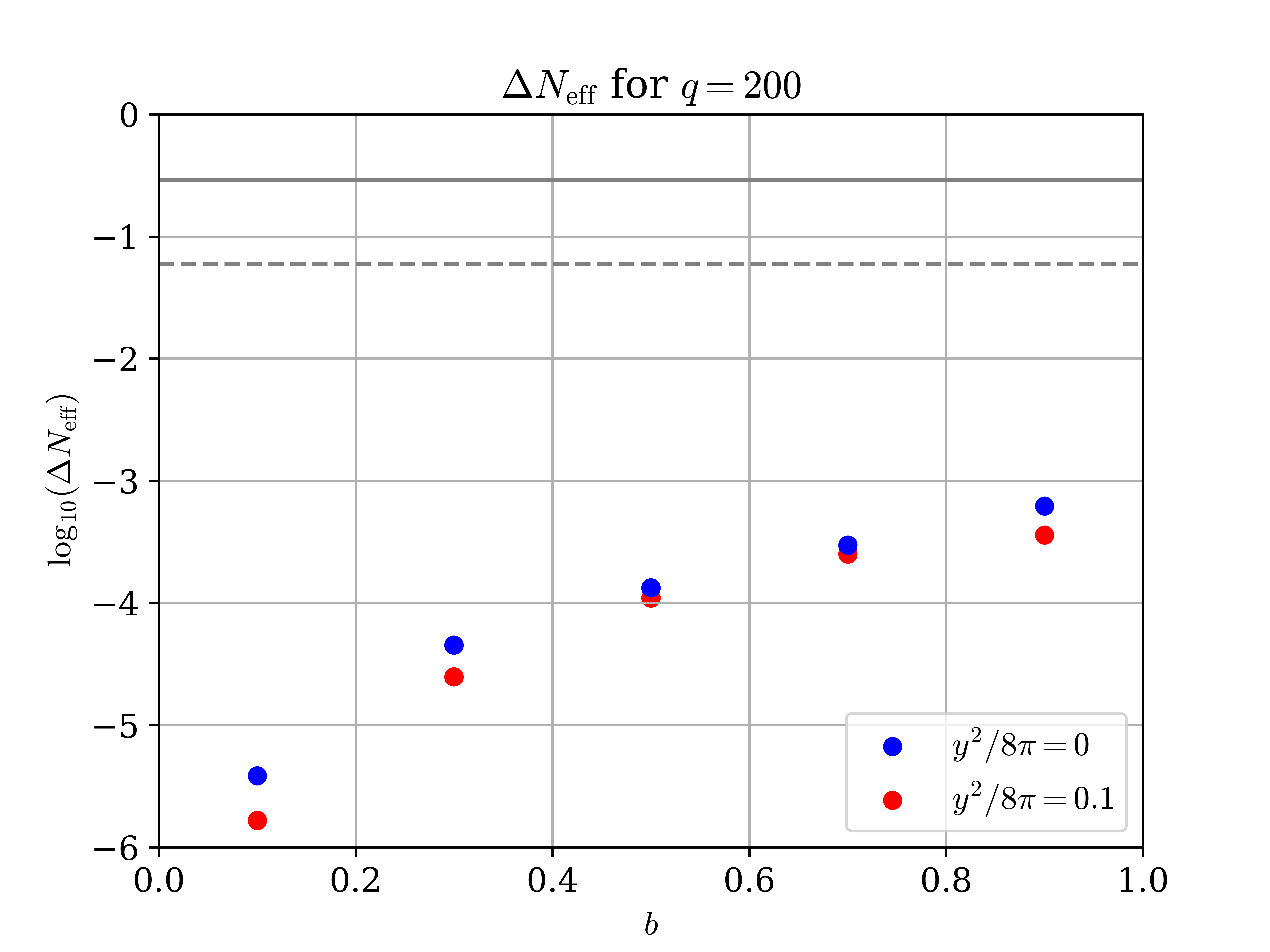}
    \caption{$\Delta N_{\text{eff}}$ against $b$ for $y=0$ (blue dots) and $y^2/(8\pi) = 0.1$, fixing $q=200$. The solid gray line denotes the present-day Planck bound, and the dashed line denotes the projected future CMB-S4 bound.}
    \label{fig:neff2}
\end{figure}

Using Eq.~\eqref{eq:NeffGW}, we could compute the GW contribution to $\Delta N_{\rm eff}$. In \Fig{fig:neff2}, we fix $q=200$ and displays the increase of $\Delta N_{\rm eff}$ with $b$. Note that the the spillway mechanism slightly reduces $\Delta N_{\text{eff}}$ for a given $b$, corresponding to a decrease in the produced total GW energy. Even when preheating mechanism maximizes the production, the contribution to $\Delta N_{\rm eff}$ is still one order of magnitude below the future CMB-S4 sensitivity.

In summary, the GWs generated in the spillway scenarios have the following features, compared to the canonical tachyonic resonance scenario:
\begin{itemize}
    \item The spectra usually possess sharper peaks, with the power falling off more quickly at higher frequencies; 
    \item The peak amplitudes are approximately similar, when both mechanisms are effective in particle production ($q \gg 10$); 
    \item The contribution to $\Delta N_{\rm eff}$ is a bit smaller, suggesting a moderately smaller amount of produced GWs. 
\end{itemize}

%%%%%%%%%%%%%%%%%%%%%%%%%%%%%%%%%%%%%%%%%
\section{Conclusions}
\label{sec:conclusions}
%%%%%%%%%%%%%%%%%%%%%%%%%%%%%%%%%%%%%%%%%
In this article, we have used classical lattice simulations to scan the parameter space of the spillway preheating model, which is tachyonic preheating assisted with a spillway --- the perturbative decay of the direct daughter particles. 
For a sufficiently large hierarchy between the masses of the inflaton and its direct daughter particles (i.e., a large $q$ parameter), the spillway induces a more radiation-like equation of state compared to its tachyonic counterpart. This manifests observationally as a moderate increase to the scalar tilt $n_s$ in the fits to inflationary observables. Conversely, at small $q$'s, the spillway halts particle production entirely, preventing efficient energy transfer out of the inflaton condensate. We also simulate the GW spectrum produced. The spillway model produces a sharper peak in the spectrum while dampening the production at higher frequencies, in comparison with the GW spectrum generated in the canonical tachyonic preheating scenario. Consequently, its contribution to the number of effective degrees of freedom beyond the standard model decreases slightly. 

Spillway preheating serves as an example of the rich dynamics in the primordial dark age, which calls for more studies. In our studies, we have to extrapolate the equation of state beyond the simulation time analytically based on some assumptions. It would be desirable to validate these assumptions numerically by improving the simulations to extend the time interval with reasonable computational resources. Beyond the primordial dark age, the setup we study could happen at the end of an early matter domination epoch, which leads to similar phenomenology, namely, modifications to $n_s$ and $r$ fit and GW production. Yet the quantitative features would be different. For example, the produced GWs would peak at a lower frequency (though still higher than the range probed by the current GW detectors), which may be more feasible for the high-frequency GW detection under development.

%%%%%%%%%%%%%%%%%%%%%%%%%%%%%%%%%%%%%%%%%
\section*{Acknowledgments}
GM is supported by the Karen T. Romer Undergraduate Teaching and Research Awards (UTRAs) at Brown. JF is supported by the NASA grant 80NSSC22K081 and the DOE grant DE-SC-0010010. QL is supported by the DOE grant DE-SC0013607, the Alfred
P. Sloan Foundation Grant No. G-2019-12504, the NSF grant PHY-2210498 and PHY-2207584, and by the Simons Foundation. This work was performed in part at Aspen Center for Physics, which is supported by National Science Foundation grant PHY-2210452. This research was conducted using computational resources and services at the Center for Computation and Visualization, Brown University.

\bibliography{ref}

\providecommand{\href}[2]{#2}\begingroup\raggedright\begin{thebibliography}{10}

\bibitem{Abbott:1982hn}
L.~Abbott, E.~Farhi, and M.~B. Wise, ``{Particle Production in the New
  Inflationary Cosmology},''
  \href{http://dx.doi.org/10.1016/0370-2693(82)90867-X}{{\em Phys. Lett. B}
  {\bfseries 117} (1982) 29}.

\bibitem{Dolgov:1982th}
A.~Dolgov and A.~D. Linde, ``{Baryon Asymmetry in Inflationary Universe},''
  \href{http://dx.doi.org/10.1016/0370-2693(82)90292-1}{{\em Phys. Lett. B}
  {\bfseries 116} (1982) 329}.

\bibitem{Albrecht:1982mp}
A.~Albrecht, P.~J. Steinhardt, M.~S. Turner, and F.~Wilczek, ``{Reheating an
  Inflationary Universe},''
  \href{http://dx.doi.org/10.1103/PhysRevLett.48.1437}{{\em Phys. Rev. Lett.}
  {\bfseries 48} (1982) 1437}.

\bibitem{Traschen:1990sw}
J.~H. Traschen and R.~H. Brandenberger, ``{Particle Production During
  Out-of-equilibrium Phase Transitions},''
  \href{http://dx.doi.org/10.1103/PhysRevD.42.2491}{{\em Phys. Rev. D}
  {\bfseries 42} (1990) 2491--2504}.

\bibitem{Dolgov:1989us}
A.~Dolgov and D.~Kirilova, ``{ON PARTICLE CREATION BY A TIME DEPENDENT SCALAR
  FIELD},'' {\em Sov. J. Nucl. Phys.} {\bfseries 51} (1990) 172--177.

\bibitem{Shtanov:1994ce}
Y.~Shtanov, J.~H. Traschen, and R.~H. Brandenberger, ``{Universe reheating
  after inflation},'' \href{http://dx.doi.org/10.1103/PhysRevD.51.5438}{{\em
  Phys. Rev. D} {\bfseries 51} (1995) 5438--5455},
  \href{http://arxiv.org/abs/hep-ph/9407247}{{\ttfamily arXiv:hep-ph/9407247}}.

\bibitem{Kofman:1994rk}
L.~Kofman, A.~D. Linde, and A.~A. Starobinsky, ``{Reheating after inflation},''
  \href{http://dx.doi.org/10.1103/PhysRevLett.73.3195}{{\em Phys. Rev. Lett.}
  {\bfseries 73} (1994) 3195--3198},
  \href{http://arxiv.org/abs/hep-th/9405187}{{\ttfamily arXiv:hep-th/9405187}}.

\bibitem{Boyanovsky:1995ud}
D.~Boyanovsky, M.~D'Attanasio, H.~de~Vega, R.~Holman, D.-S. Lee, and A.~Singh,
  ``{Reheating the postinflationary universe},''
  \href{http://arxiv.org/abs/hep-ph/9505220}{{\ttfamily arXiv:hep-ph/9505220}}.

\bibitem{Yoshimura:1995gc}
M.~Yoshimura, ``{Catastrophic particle production under periodic
  perturbation},'' \href{http://dx.doi.org/10.1143/PTP.94.873}{{\em Prog.
  Theor. Phys.} {\bfseries 94} (1995) 873--898},
  \href{http://arxiv.org/abs/hep-th/9506176}{{\ttfamily arXiv:hep-th/9506176}}.

\bibitem{Kaiser:1995fb}
D.~I. Kaiser, ``{Post inflation reheating in an expanding universe},''
  \href{http://dx.doi.org/10.1103/PhysRevD.53.1776}{{\em Phys. Rev. D}
  {\bfseries 53} (1996) 1776--1783},
  \href{http://arxiv.org/abs/astro-ph/9507108}{{\ttfamily
  arXiv:astro-ph/9507108}}.

\bibitem{Kofman:1997yn}
L.~Kofman, A.~D. Linde, and A.~A. Starobinsky, ``{Towards the theory of
  reheating after inflation},''
  \href{http://dx.doi.org/10.1103/PhysRevD.56.3258}{{\em Phys. Rev. D}
  {\bfseries 56} (1997) 3258--3295},
  \href{http://arxiv.org/abs/hep-ph/9704452}{{\ttfamily arXiv:hep-ph/9704452}}.

\bibitem{Allahverdi:2010xz}
R.~Allahverdi, R.~Brandenberger, F.-Y. Cyr-Racine, and A.~Mazumdar,
  ``{Reheating in Inflationary Cosmology: Theory and Applications},''
  \href{http://dx.doi.org/10.1146/annurev.nucl.012809.104511}{{\em Ann. Rev.
  Nucl. Part. Sci.} {\bfseries 60} (2010) 27--51},
  \href{http://arxiv.org/abs/1001.2600}{{\ttfamily arXiv:1001.2600 [hep-th]}}.

\bibitem{Amin:2014eta}
M.~A. Amin, M.~P. Hertzberg, D.~I. Kaiser, and J.~Karouby, ``{Nonperturbative
  Dynamics Of Reheating After Inflation: A Review},''
  \href{http://dx.doi.org/10.1142/S0218271815300037}{{\em Int. J. Mod. Phys. D}
  {\bfseries 24} (2014) 1530003},
  \href{http://arxiv.org/abs/1410.3808}{{\ttfamily arXiv:1410.3808 [hep-ph]}}.

\bibitem{Dufaux:2006ee}
J.~F. Dufaux, G.~N. Felder, L.~Kofman, M.~Peloso, and D.~Podolsky,
  ``{Preheating with trilinear interactions: Tachyonic resonance},''
  \href{http://dx.doi.org/10.1088/1475-7516/2006/07/006}{{\em JCAP} {\bfseries
  07} (2006) 006}, \href{http://arxiv.org/abs/hep-ph/0602144}{{\ttfamily
  arXiv:hep-ph/0602144}}.

\bibitem{Deskins:2013dwa}
J.~Deskins, J.~T. Giblin, and R.~R. Caldwell, ``{Gauge Field Preheating at the
  End of Inflation},'' \href{http://dx.doi.org/10.1103/PhysRevD.88.063530}{{\em
  Phys. Rev. D} {\bfseries 88} no.~6, (2013) 063530},
  \href{http://arxiv.org/abs/1305.7226}{{\ttfamily arXiv:1305.7226
  [astro-ph.CO]}}.

\bibitem{Adshead:2015pva}
P.~Adshead, J.~T. Giblin, T.~R. Scully, and E.~I. Sfakianakis,
  ``{Gauge-preheating and the end of axion inflation},''
  \href{http://dx.doi.org/10.1088/1475-7516/2015/12/034}{{\em JCAP} {\bfseries
  12} (2015) 034}, \href{http://arxiv.org/abs/1502.06506}{{\ttfamily
  arXiv:1502.06506 [astro-ph.CO]}}.

\bibitem{Adshead:2017xll}
P.~Adshead, J.~T. Giblin, and Z.~J. Weiner, ``{Non-Abelian gauge preheating},''
  \href{http://dx.doi.org/10.1103/PhysRevD.96.123512}{{\em Phys. Rev. D}
  {\bfseries 96} no.~12, (2017) 123512},
  \href{http://arxiv.org/abs/1708.02944}{{\ttfamily arXiv:1708.02944
  [hep-ph]}}.

\bibitem{Cuissa:2018oiw}
J.~R.~C. Cuissa and D.~G. Figueroa, ``{Lattice formulation of axion inflation.
  Application to preheating},''
  \href{http://dx.doi.org/10.1088/1475-7516/2019/06/002}{{\em JCAP} {\bfseries
  06} (2019) 002}, \href{http://arxiv.org/abs/1812.03132}{{\ttfamily
  arXiv:1812.03132 [astro-ph.CO]}}.

\bibitem{Fan:2021otj}
J.~Fan, K.~D. Lozanov, and Q.~Lu, ``{Spillway Preheating},''
  \href{http://dx.doi.org/10.1007/JHEP05(2021)069}{{\em JHEP} {\bfseries 05}
  (2021) 069}, \href{http://arxiv.org/abs/2101.11008}{{\ttfamily
  arXiv:2101.11008 [hep-ph]}}.

\bibitem{Felder:1998vq}
G.~N. Felder, L.~Kofman, and A.~D. Linde, ``{Instant preheating},''
  \href{http://dx.doi.org/10.1103/PhysRevD.59.123523}{{\em Phys. Rev. D}
  {\bfseries 59} (1999) 123523},
  \href{http://arxiv.org/abs/hep-ph/9812289}{{\ttfamily arXiv:hep-ph/9812289}}.

\bibitem{Garcia-Bellido:2008ycs}
J.~Garcia-Bellido, D.~G. Figueroa, and J.~Rubio, ``{Preheating in the Standard
  Model with the Higgs-Inflaton coupled to gravity},''
  \href{http://dx.doi.org/10.1103/PhysRevD.79.063531}{{\em Phys. Rev. D}
  {\bfseries 79} (2009) 063531},
  \href{http://arxiv.org/abs/0812.4624}{{\ttfamily arXiv:0812.4624 [hep-ph]}}.

\bibitem{Repond:2016sol}
J.~Repond and J.~Rubio, ``{Combined Preheating on the lattice with applications
  to Higgs inflation},''
  \href{http://dx.doi.org/10.1088/1475-7516/2016/07/043}{{\em JCAP} {\bfseries
  07} (2016) 043}, \href{http://arxiv.org/abs/1604.08238}{{\ttfamily
  arXiv:1604.08238 [astro-ph.CO]}}.

\bibitem{Kasuya:1996np}
S.~Kasuya and M.~Kawasaki, ``{Restriction to parametric resonant decay after
  inflation},'' \href{http://dx.doi.org/10.1016/S0370-2693(96)01216-6}{{\em
  Phys. Lett. B} {\bfseries 388} (1996) 686--691},
  \href{http://arxiv.org/abs/hep-ph/9603317}{{\ttfamily arXiv:hep-ph/9603317}}.

\bibitem{Bezrukov:2008ut}
F.~Bezrukov, D.~Gorbunov, and M.~Shaposhnikov, ``{On initial conditions for the
  Hot Big Bang},'' \href{http://dx.doi.org/10.1088/1475-7516/2009/06/029}{{\em
  JCAP} {\bfseries 06} (2009) 029},
  \href{http://arxiv.org/abs/0812.3622}{{\ttfamily arXiv:0812.3622 [hep-ph]}}.

\bibitem{Mukaida:2012bz}
K.~Mukaida and K.~Nakayama, ``{Dissipative Effects on Reheating after
  Inflation},'' \href{http://dx.doi.org/10.1088/1475-7516/2013/03/002}{{\em
  JCAP} {\bfseries 03} (2013) 002},
  \href{http://arxiv.org/abs/1212.4985}{{\ttfamily arXiv:1212.4985 [hep-ph]}}.

\bibitem{Kost:2021rbi}
J.~Kost, C.~S. Shin, and T.~Terada, ``{Massless preheating and electroweak
  vacuum metastability},''
  \href{http://dx.doi.org/10.1103/PhysRevD.105.043508}{{\em Phys. Rev. D}
  {\bfseries 105} no.~4, (2022) 043508},
  \href{http://arxiv.org/abs/2105.06939}{{\ttfamily arXiv:2105.06939
  [hep-ph]}}.

\bibitem{Garcia:2021iag}
M.~A.~G. Garcia, K.~Kaneta, Y.~Mambrini, K.~A. Olive, and S.~Verner,
  ``{Freeze-in from preheating},''
  \href{http://dx.doi.org/10.1088/1475-7516/2022/03/016}{{\em JCAP} {\bfseries
  03} no.~03, (2022) 016}, \href{http://arxiv.org/abs/2109.13280}{{\ttfamily
  arXiv:2109.13280 [hep-ph]}}.

\bibitem{Liddle:2003as}
A.~R. Liddle and S.~M. Leach, ``{How long before the end of inflation were
  observable perturbations produced?},''
  \href{http://dx.doi.org/10.1103/PhysRevD.68.103503}{{\em Phys. Rev. D}
  {\bfseries 68} (2003) 103503},
  \href{http://arxiv.org/abs/astro-ph/0305263}{{\ttfamily
  arXiv:astro-ph/0305263}}.

\bibitem{Dai:2014jja}
L.~Dai, M.~Kamionkowski, and J.~Wang, ``{Reheating constraints to inflationary
  models},'' \href{http://dx.doi.org/10.1103/PhysRevLett.113.041302}{{\em Phys.
  Rev. Lett.} {\bfseries 113} (2014) 041302},
  \href{http://arxiv.org/abs/1404.6704}{{\ttfamily arXiv:1404.6704
  [astro-ph.CO]}}.

\bibitem{Munoz:2014eqa}
J.~B. Munoz and M.~Kamionkowski, ``{Equation-of-State Parameter for
  Reheating},'' \href{http://dx.doi.org/10.1103/PhysRevD.91.043521}{{\em Phys.
  Rev. D} {\bfseries 91} no.~4, (2015) 043521},
  \href{http://arxiv.org/abs/1412.0656}{{\ttfamily arXiv:1412.0656
  [astro-ph.CO]}}.

\bibitem{Martin:2016oyk}
J.~Martin, C.~Ringeval, and V.~Vennin, ``{Information Gain on Reheating: the
  One Bit Milestone},''
  \href{http://dx.doi.org/10.1103/PhysRevD.93.103532}{{\em Phys. Rev. D}
  {\bfseries 93} no.~10, (2016) 103532},
  \href{http://arxiv.org/abs/1603.02606}{{\ttfamily arXiv:1603.02606
  [astro-ph.CO]}}.

\bibitem{Hardwick:2016whe}
R.~J. Hardwick, V.~Vennin, K.~Koyama, and D.~Wands, ``{Constraining Curvatonic
  Reheating},'' \href{http://dx.doi.org/10.1088/1475-7516/2016/08/042}{{\em
  JCAP} {\bfseries 08} (2016) 042},
  \href{http://arxiv.org/abs/1606.01223}{{\ttfamily arXiv:1606.01223
  [astro-ph.CO]}}.

\bibitem{Lozanov:2016hid}
K.~D. Lozanov and M.~A. Amin, ``{Equation of State and Duration to Radiation
  Domination after Inflation},''
  \href{http://dx.doi.org/10.1103/PhysRevLett.119.061301}{{\em Phys. Rev.
  Lett.} {\bfseries 119} no.~6, (2017) 061301},
  \href{http://arxiv.org/abs/1608.01213}{{\ttfamily arXiv:1608.01213
  [astro-ph.CO]}}.

\bibitem{Lozanov:2017hjm}
K.~D. Lozanov and M.~A. Amin, ``{Self-resonance after inflation: oscillons,
  transients and radiation domination},''
  \href{http://dx.doi.org/10.1103/PhysRevD.97.023533}{{\em Phys. Rev. D}
  {\bfseries 97} no.~2, (2018) 023533},
  \href{http://arxiv.org/abs/1710.06851}{{\ttfamily arXiv:1710.06851
  [astro-ph.CO]}}.

\bibitem{Antusch:2020iyq}
S.~Antusch, D.~G. Figueroa, K.~Marschall, and F.~Torrenti, ``{Energy
  distribution and equation of state of the early Universe: matching the end of
  inflation and the onset of radiation domination},''
  \href{http://dx.doi.org/10.1016/j.physletb.2020.135888}{{\em Phys. Lett. B}
  {\bfseries 811} (2020) 135888},
  \href{http://arxiv.org/abs/2005.07563}{{\ttfamily arXiv:2005.07563
  [astro-ph.CO]}}.

\bibitem{Bettoni:2021zhq}
D.~Bettoni, A.~Lopez-Eiguren, and J.~Rubio, ``{Hubble-induced phase transitions
  on the lattice with applications to Ricci reheating},''
  \href{http://dx.doi.org/10.1088/1475-7516/2022/01/002}{{\em JCAP} {\bfseries
  01} no.~01, (2022) 002}, \href{http://arxiv.org/abs/2107.09671}{{\ttfamily
  arXiv:2107.09671 [hep-ph]}}.

\bibitem{Antusch:2022mqv}
S.~Antusch, K.~Marschall, and F.~Torrenti, ``{Characterizing the
  post-inflationary reheating history. Part II. Multiple interacting daughter
  fields},'' \href{http://dx.doi.org/10.1088/1475-7516/2023/02/019}{{\em JCAP}
  {\bfseries 02} (2023) 019}, \href{http://arxiv.org/abs/2206.06319}{{\ttfamily
  arXiv:2206.06319 [astro-ph.CO]}}.

\bibitem{Lodman:2023yrc}
J.~Lodman, Q.~Lu, and L.~Randall, ``{Savior Curvatons and Large
  non-Gaussianity},'' \href{http://arxiv.org/abs/2306.13128}{{\ttfamily
  arXiv:2306.13128 [astro-ph.CO]}}.

\bibitem{Barman:2023opy}
B.~Barman, N.~Bernal, and J.~Rubio, ``{Rescuing Gravitational-Reheating in
  Chaotic Inflation},'' \href{http://arxiv.org/abs/2310.06039}{{\ttfamily
  arXiv:2310.06039 [hep-ph]}}.

\bibitem{Khlebnikov:1997di}
S.~Khlebnikov and I.~Tkachev, ``{Relic gravitational waves produced after
  preheating},'' \href{http://dx.doi.org/10.1103/PhysRevD.56.653}{{\em Phys.
  Rev. D} {\bfseries 56} (1997) 653--660},
  \href{http://arxiv.org/abs/hep-ph/9701423}{{\ttfamily arXiv:hep-ph/9701423}}.

\bibitem{Easther:2006vd}
R.~Easther, J.~Giblin, John~T., and E.~A. Lim, ``{Gravitational Wave Production
  At The End Of Inflation},''
  \href{http://dx.doi.org/10.1103/PhysRevLett.99.221301}{{\em Phys. Rev. Lett.}
  {\bfseries 99} (2007) 221301},
  \href{http://arxiv.org/abs/astro-ph/0612294}{{\ttfamily
  arXiv:astro-ph/0612294}}.

\bibitem{Easther:2006gt}
R.~Easther and E.~A. Lim, ``{Stochastic gravitational wave production after
  inflation},'' \href{http://dx.doi.org/10.1088/1475-7516/2006/04/010}{{\em
  JCAP} {\bfseries 04} (2006) 010},
  \href{http://arxiv.org/abs/astro-ph/0601617}{{\ttfamily
  arXiv:astro-ph/0601617}}.

\bibitem{GarciaBellido:2007af}
J.~Garcia-Bellido, D.~G. Figueroa, and A.~Sastre, ``{A Gravitational Wave
  Background from Reheating after Hybrid Inflation},''
  \href{http://dx.doi.org/10.1103/PhysRevD.77.043517}{{\em Phys. Rev. D}
  {\bfseries 77} (2008) 043517},
  \href{http://arxiv.org/abs/0707.0839}{{\ttfamily arXiv:0707.0839 [hep-ph]}}.

\bibitem{Dufaux:2007pt}
J.~F. Dufaux, A.~Bergman, G.~N. Felder, L.~Kofman, and J.-P. Uzan, ``{Theory
  and Numerics of Gravitational Waves from Preheating after Inflation},''
  \href{http://dx.doi.org/10.1103/PhysRevD.76.123517}{{\em Phys. Rev. D}
  {\bfseries 76} (2007) 123517},
  \href{http://arxiv.org/abs/0707.0875}{{\ttfamily arXiv:0707.0875
  [astro-ph]}}.

\bibitem{Dufaux:2008dn}
J.-F. Dufaux, G.~Felder, L.~Kofman, and O.~Navros, ``{Gravity Waves from
  Tachyonic Preheating after Hybrid Inflation},''
  \href{http://dx.doi.org/10.1088/1475-7516/2009/03/001}{{\em JCAP} {\bfseries
  03} (2009) 001}, \href{http://arxiv.org/abs/0812.2917}{{\ttfamily
  arXiv:0812.2917 [astro-ph]}}.

\bibitem{Dufaux:2010cf}
J.-F. Dufaux, D.~G. Figueroa, and J.~Garcia-Bellido, ``{Gravitational Waves
  from Abelian Gauge Fields and Cosmic Strings at Preheating},''
  \href{http://dx.doi.org/10.1103/PhysRevD.82.083518}{{\em Phys. Rev. D}
  {\bfseries 82} (2010) 083518},
  \href{http://arxiv.org/abs/1006.0217}{{\ttfamily arXiv:1006.0217
  [astro-ph.CO]}}.

\bibitem{Bethke:2013vca}
L.~Bethke, D.~G. Figueroa, and A.~Rajantie, ``{On the Anisotropy of the
  Gravitational Wave Background from Massless Preheating},''
  \href{http://dx.doi.org/10.1088/1475-7516/2014/06/047}{{\em JCAP} {\bfseries
  06} (2014) 047}, \href{http://arxiv.org/abs/1309.1148}{{\ttfamily
  arXiv:1309.1148 [astro-ph.CO]}}.

\bibitem{Adshead:2018doq}
P.~Adshead, J.~T. Giblin, and Z.~J. Weiner, ``{Gravitational waves from gauge
  preheating},'' \href{http://dx.doi.org/10.1103/PhysRevD.98.043525}{{\em Phys.
  Rev. D} {\bfseries 98} no.~4, (2018) 043525},
  \href{http://arxiv.org/abs/1805.04550}{{\ttfamily arXiv:1805.04550
  [astro-ph.CO]}}.

\bibitem{Kitajima:2018zco}
N.~Kitajima, J.~Soda, and Y.~Urakawa, ``{Gravitational wave forest from string
  axiverse},'' \href{http://dx.doi.org/10.1088/1475-7516/2018/10/008}{{\em
  JCAP} {\bfseries 10} (2018) 008},
  \href{http://arxiv.org/abs/1807.07037}{{\ttfamily arXiv:1807.07037
  [astro-ph.CO]}}.

\bibitem{Bartolo:2016ami}
N.~Bartolo {\em et~al.}, ``{Science with the space-based interferometer LISA.
  IV: Probing inflation with gravitational waves},''
  \href{http://dx.doi.org/10.1088/1475-7516/2016/12/026}{{\em JCAP} {\bfseries
  12} (2016) 026}, \href{http://arxiv.org/abs/1610.06481}{{\ttfamily
  arXiv:1610.06481 [astro-ph.CO]}}.

\bibitem{Figueroa:2017vfa}
D.~G. Figueroa and F.~Torrenti, ``{Gravitational wave production from
  preheating: parameter dependence},''
  \href{http://dx.doi.org/10.1088/1475-7516/2017/10/057}{{\em JCAP} {\bfseries
  10} (2017) 057}, \href{http://arxiv.org/abs/1707.04533}{{\ttfamily
  arXiv:1707.04533 [astro-ph.CO]}}.

\bibitem{Caprini:2018mtu}
C.~Caprini and D.~G. Figueroa, ``{Cosmological Backgrounds of Gravitational
  Waves},'' \href{http://dx.doi.org/10.1088/1361-6382/aac608}{{\em Class.
  Quant. Grav.} {\bfseries 35} no.~16, (2018) 163001},
  \href{http://arxiv.org/abs/1801.04268}{{\ttfamily arXiv:1801.04268
  [astro-ph.CO]}}.

\bibitem{Bartolo:2018qqn}
N.~Bartolo, V.~Domcke, D.~G. Figueroa, J.~Garc\'\i{}a-Bellido, M.~Peloso,
  M.~Pieroni, A.~Ricciardone, M.~Sakellariadou, L.~Sorbo, and G.~Tasinato,
  ``{Probing non-Gaussian Stochastic Gravitational Wave Backgrounds with
  LISA},'' \href{http://dx.doi.org/10.1088/1475-7516/2018/11/034}{{\em JCAP}
  {\bfseries 11} (2018) 034}, \href{http://arxiv.org/abs/1806.02819}{{\ttfamily
  arXiv:1806.02819 [astro-ph.CO]}}.

\bibitem{Lozanov:2019ylm}
K.~D. Lozanov and M.~A. Amin, ``{Gravitational perturbations from oscillons and
  transients after inflation},''
  \href{http://dx.doi.org/10.1103/PhysRevD.99.123504}{{\em Phys. Rev. D}
  {\bfseries 99} no.~12, (2019) 123504},
  \href{http://arxiv.org/abs/1902.06736}{{\ttfamily arXiv:1902.06736
  [astro-ph.CO]}}.

\bibitem{Adshead:2019igv}
P.~Adshead, J.~T. Giblin, M.~Pieroni, and Z.~J. Weiner, ``{Constraining Axion
  Inflation with Gravitational Waves across 29 Decades in Frequency},''
  \href{http://dx.doi.org/10.1103/PhysRevLett.124.171301}{{\em Phys. Rev.
  Lett.} {\bfseries 124} no.~17, (2020) 171301},
  \href{http://arxiv.org/abs/1909.12843}{{\ttfamily arXiv:1909.12843
  [astro-ph.CO]}}.

\bibitem{Adshead:2019lbr}
P.~Adshead, J.~T. Giblin, M.~Pieroni, and Z.~J. Weiner, ``{Constraining axion
  inflation with gravitational waves from preheating},''
  \href{http://dx.doi.org/10.1103/PhysRevD.101.083534}{{\em Phys. Rev. D}
  {\bfseries 101} no.~8, (2020) 083534},
  \href{http://arxiv.org/abs/1909.12842}{{\ttfamily arXiv:1909.12842
  [astro-ph.CO]}}.

\bibitem{Amin:2018kkg}
M.~A. Amin, J.~Fan, K.~D. Lozanov, and M.~Reece, ``{Cosmological dynamics of
  Higgs potential fine tuning},''
  \href{http://dx.doi.org/10.1103/PhysRevD.99.035008}{{\em Phys. Rev. D}
  {\bfseries 99} no.~3, (2019) 035008},
  \href{http://arxiv.org/abs/1802.00444}{{\ttfamily arXiv:1802.00444
  [hep-ph]}}.

\bibitem{Akrami:2018odb}
{\bfseries Planck} Collaboration, Y.~Akrami {\em et~al.}, ``{Planck 2018
  results. X. Constraints on inflation},''
  \href{http://dx.doi.org/10.1051/0004-6361/201833887}{{\em Astron. Astrophys.}
  {\bfseries 641} (2020) A10},
  \href{http://arxiv.org/abs/1807.06211}{{\ttfamily arXiv:1807.06211
  [astro-ph.CO]}}.

\bibitem{Felder:2000hq}
G.~N. Felder and I.~Tkachev, ``{LATTICEEASY: A Program for lattice simulations
  of scalar fields in an expanding universe},''
  \href{http://dx.doi.org/10.1016/j.cpc.2008.02.009}{{\em Comput. Phys.
  Commun.} {\bfseries 178} (2008) 929--932},
  \href{http://arxiv.org/abs/hep-ph/0011159}{{\ttfamily arXiv:hep-ph/0011159}}.

\bibitem{nsr2}
{\bfseries CMB-S4} Collaboration, K.~Abazajian {\em et~al.}, ``{CMB-S4:
  Forecasting Constraints on Primordial Gravitational Waves},''
  \href{http://dx.doi.org/10.3847/1538-4357/ac1596}{{\em Astrophys. J.}
  {\bfseries 926} no.~1, (2022) 54},
  \href{http://arxiv.org/abs/2008.12619}{{\ttfamily arXiv:2008.12619
  [astro-ph.CO]}}.

\bibitem{nsr}
K.~Abazajian {\em et~al.}, ``{CMB-S4 Science Case, Reference Design, and
  Project Plan},'' \href{http://arxiv.org/abs/1907.04473}{{\ttfamily
  arXiv:1907.04473 [astro-ph.IM]}}.

\bibitem{LINDE1983177}
A.~D. Linde, ``{Chaotic Inflation},''
  \href{http://dx.doi.org/10.1016/0370-2693(83)90837-7}{{\em Phys. Lett. B}
  {\bfseries 129} (1983) 177--181}.

\bibitem{Belinsky:1985zd}
V.~A. Belinsky, I.~M. Khalatnikov, L.~P. Grishchuk, and Y.~B. Zeldovich,
  ``{INFLATIONARY STAGES IN COSMOLOGICAL MODELS WITH A SCALAR FIELD},''
  \href{http://dx.doi.org/10.1016/0370-2693(85)90644-6}{{\em Phys. Lett. B}
  {\bfseries 155} (1985) 232--236}.

\bibitem{PIRAN1985331}
T.~Piran and R.~M. Williams, ``Inflation in universes with a massive scalar
  field,''
  \href{http://dx.doi.org/https://doi.org/10.1016/0370-2693(85)90291-6}{{\em
  Physics Letters B} {\bfseries 163} no.~5, (1985) 331--335}.

\bibitem{McAllister_2010}
L.~McAllister, E.~Silverstein, and A.~Westphal, ``{Gravity Waves and Linear
  Inflation from Axion Monodromy},''
  \href{http://dx.doi.org/10.1103/PhysRevD.82.046003}{{\em Phys. Rev. D}
  {\bfseries 82} (2010) 046003},
  \href{http://arxiv.org/abs/0808.0706}{{\ttfamily arXiv:0808.0706 [hep-th]}}.

\bibitem{Silverstein_2008}
E.~Silverstein and A.~Westphal, ``{Monodromy in the CMB: Gravity Waves and
  String Inflation},'' \href{http://dx.doi.org/10.1103/PhysRevD.78.106003}{{\em
  Phys. Rev. D} {\bfseries 78} (2008) 106003},
  \href{http://arxiv.org/abs/0803.3085}{{\ttfamily arXiv:0803.3085 [hep-th]}}.

\bibitem{Boubekeur_2005}
L.~Boubekeur and D.~H. Lyth, ``{Hilltop inflation},''
  \href{http://dx.doi.org/10.1088/1475-7516/2005/07/010}{{\em JCAP} {\bfseries
  07} (2005) 010}, \href{http://arxiv.org/abs/hep-ph/0502047}{{\ttfamily
  arXiv:hep-ph/0502047}}.

\bibitem{Germ_n_2021}
G.~German, ``{Quartic hilltop inflation revisited},''
  \href{http://dx.doi.org/10.1088/1475-7516/2021/02/034}{{\em JCAP} {\bfseries
  02} (2021) 034}, \href{http://arxiv.org/abs/2011.12804}{{\ttfamily
  arXiv:2011.12804 [astro-ph.CO]}}.

\bibitem{Kallosh:2013hoa}
R.~Kallosh and A.~Linde, ``{Universality Class in Conformal Inflation},''
  \href{http://dx.doi.org/10.1088/1475-7516/2013/07/002}{{\em JCAP} {\bfseries
  07} (2013) 002}, \href{http://arxiv.org/abs/1306.5220}{{\ttfamily
  arXiv:1306.5220 [hep-th]}}.

\bibitem{Lozanov:2019jxc}
K.~D. Lozanov, ``{Lectures on Reheating after Inflation},''
  \href{http://arxiv.org/abs/1907.04402}{{\ttfamily arXiv:1907.04402
  [astro-ph.CO]}}.

\bibitem{PhysRevD.77.043517}
J.~Garcia-Bellido, D.~G. Figueroa, and A.~Sastre, ``{A Gravitational Wave
  Background from Reheating after Hybrid Inflation},''
  \href{http://dx.doi.org/10.1103/PhysRevD.77.043517}{{\em Phys. Rev. D}
  {\bfseries 77} (2008) 043517},
  \href{http://arxiv.org/abs/0707.0839}{{\ttfamily arXiv:0707.0839 [hep-ph]}}.

\bibitem{Workman:2022ynf}
{\bfseries Particle Data Group} Collaboration, R.~L. Workman {\em et~al.},
  ``{Review of Particle Physics},''
  \href{http://dx.doi.org/10.1093/ptep/ptac097}{{\em PTEP} {\bfseries 2022}
  (2022) 083C01}.

\bibitem{Figueroa_2023}
D.~G. Figueroa, A.~Florio, F.~Torrenti, and W.~Valkenburg, ``{CosmoLattice: A
  modern code for lattice simulations of scalar and gauge field dynamics in an
  expanding universe},''
  \href{http://dx.doi.org/10.1016/j.cpc.2022.108586}{{\em Comput. Phys.
  Commun.} {\bfseries 283} (2023) 108586},
  \href{http://arxiv.org/abs/2102.01031}{{\ttfamily arXiv:2102.01031
  [astro-ph.CO]}}.

\end{thebibliography}\endgroup
\bibliographystyle{utphys}
\end{document}